\documentclass[prx,aps,twocolumn,nopacs,superscriptaddress,nofootinbib,longbibliography]{revtex4-1}

\usepackage{amsmath}  \usepackage{amssymb}  \usepackage{amsfonts}  \usepackage{bm}  \usepackage{bbm}   \usepackage{bbold}  \usepackage{braket}  \usepackage{color}  \usepackage{comment}  \usepackage{dcolumn}  \usepackage{enumerate}  \usepackage{epsfig}  \usepackage{gensymb}  \usepackage{graphicx}  \usepackage{indentfirst}  \usepackage{lmodern}  \usepackage{mathrsfs}  \usepackage{mathtools}  \usepackage{psfrag}  \usepackage{pst-all}  \usepackage{soul}  
\usepackage{xcolor}
\usepackage{float}
\usepackage[colorlinks,linkcolor=blue,citecolor=blue,urlcolor=blue,hyperindex,driverfallback=dvipdfm]{hyperref}
\usepackage[T1]{fontenc}

\def\ii{{\rm i}}  \def\ee{{\rm e}}
\def\me{m_{\rm e}}  \def\kB{{k_{\rm B}}}
\def\Ree{{\rm Re}}  \def\Imm{{\rm Im}}
\def\Ab{{\bf A}}        \def\Eb{{\bf E}}  \def\eb{{\bf e}}                              \def\Qb{{\bf Q}}  \def\qb{{\bf q}}  \def\Rb{{\bf R}}  \def\rb{{\bf r}}      \def\vb{{\bf v}}
\def\xx{\hat{\bf x}}  \def\yy{\hat{\bf y}}  \def\zz{\hat{\bf z}}           
      \def\wp{\omega_{\rm p}}   
\def\Hh{\hat{\mathcal{H}}}  \def\Hint{\Hh_{\rm int}}  
\def\ah{\hat{a}}    \def\ahd{\ah^{\dagger}} 
\def\Ah{\hat{A}}
        
\def\vE{\vec{\mathcal{E}}}  \def\QQ{\hat{\bf Q}}  \def\qq{\hat{\bf q}}

\begin{document} 

\title{Free-electron decoherence: Theory and applications}
\author{Cruz~I.~Velasco} 
\affiliation{ICFO-Institut de Ciencies Fotoniques, The Barcelona Institute of Science and Technology, 08860 Castelldefels (Barcelona), Spain}
\author{Valerio~Di~Giulio} 
\affiliation{ICFO-Institut de Ciencies Fotoniques, The Barcelona Institute of Science and Technology, 08860 Castelldefels (Barcelona), Spain}
\author{F.~Javier~Garc\'{\i}a~de~Abajo} 
\email{javier.garciadeabajo@nanophotonics.es}
\affiliation{ICFO-Institut de Ciencies Fotoniques, The Barcelona Institute of Science and Technology, 08860 Castelldefels (Barcelona), Spain}
\affiliation{ICREA-Instituci\'o Catalana de Recerca i Estudis Avan\c{c}ats, Passeig Llu\'{\i}s Companys 23, 08010 Barcelona, Spain}

\begin{abstract}
Electron microscopy relies on the spatial coherence of electron beams to generate atomic-scale images using interference and diffraction, which can be degraded by inelastic scattering processes that induce decoherence. Here, we present a theoretical study of decoherence arising from the electromagnetic interaction of free electrons with bulk materials and planar surfaces. We show that bulk plasmons dominate decoherence in Al and Au, while electronic excitations above the band gap, supplemented by weaker coupling to phononic and guided modes, are the primary channels in ionic insulators such as LiF. A thermal population of electromagnetic modes leads to a divergence in the energy-loss probability at low frequencies, which in turn produces a pronounced temperature dependence. We show that this effect can be exploited for nanoscale thermometry, predicting that optimized energy-filtered holography enables $\sim0.1\%$ changes in fringe visibility for physically viable temperature variations in metals. Through these results, we establish a unified theoretical framework to describe free-electron decoherence in the bulk and surfaces of arbitrary materials.
\end{abstract}
\date{\today}

\maketitle

\section{Introduction}

In electron microscopy, free electrons serve as unique nanoscale probes thanks to their evanescent broadband electromagnetic fields and short de Broglie wavelength ($\lambda_e=2-39$~pm for kinetic energies in the $300-1$\,keV range), which pushes the diffraction limit well below the size of an atom. These properties allow free electrons to couple to modes that are inaccessible to free-space radiation while also enabling atomic-scale spatial resolution. In addition, electron microscopes currently implement techniques such as low-energy \cite{DG1927,P1974} and ultrafast \cite{WPL16,FML22} electron diffraction, as well as holography \cite{LL02,WBT18}, which can be performed using laterally extended electron beams (e-beams) and exploit the wave-like nature of free electrons to image the phase produced by elastic scattering by the atomic lattice of a specimen. In this context, the inelastic scattering associated with the exchange of energy quanta between the electron and various types of excitations (e.g., electronic \cite{APZ97,MPV03,L04,HL06,M06_3,SH07,H10_2,BZB18,KRS20}, radiative \cite{F93,F97,paper425}, and surface modes in general \cite{F93,MPV03,HL06,paper357}) can undermine this working principle by introducing decoherence between spatially separated components. 

Free-electron decoherence is produced when separated spatial components of the electron wavefunction couple to the environment with different amplitudes, thereby generating which-path information. Because incoherent components cannot interfere, they contribute only a uniform background to the interference pattern, reducing the contrast and degrading the visibility of interference fringes in electron holography. However, interference from inelastically scattered electrons remains possible when different lateral positions of the incident wavefunction interact similarly with the extended modes of a specimen. In particular, different lateral positions of energy-filtered electrons that have excited plasmons while traversing a metallic film can still interfere if the propagation length of those plasmons is larger than the separation between those positions. This is the so-called inelastic electron holography, which has been demonstrated in several experiments \cite{LF00,H05,PLV06,PVS07,VBS08}. Understanding of the different sources of decoherence thus becomes an important goal.

Earlier theoretical studies focused on the exploration of decoherence in two-path electron interferometry near planar surfaces \cite{APZ97,MPV03,L04,HL06,M06_3,H11_2,SB12}, followed by experimental demonstrations \cite{SH07,H10_2,BZB18,KRS20,CB20}. In these works, the dominant contributions to decoherence arise from coupling to electromagnetic modes propagating along the material surface with phase velocities matching the electron speed. More recently, decoherence has also been predicted in situations where electrons pass near the edge of extended structures, resulting in the emission of free radiation \cite{paper425}. In this regime, temperature-induced spontaneous emission and absorption of low-energy radiative modes drastically increase the loss of coherence. Since the electron interacts only with radiation fields without exciting internal modes of the sample, this mechanism has been proposed as a route for noncontact detection of distant objects \cite{paper425}.

\begin{figure*}
\centering\includegraphics[width=1.0\linewidth]{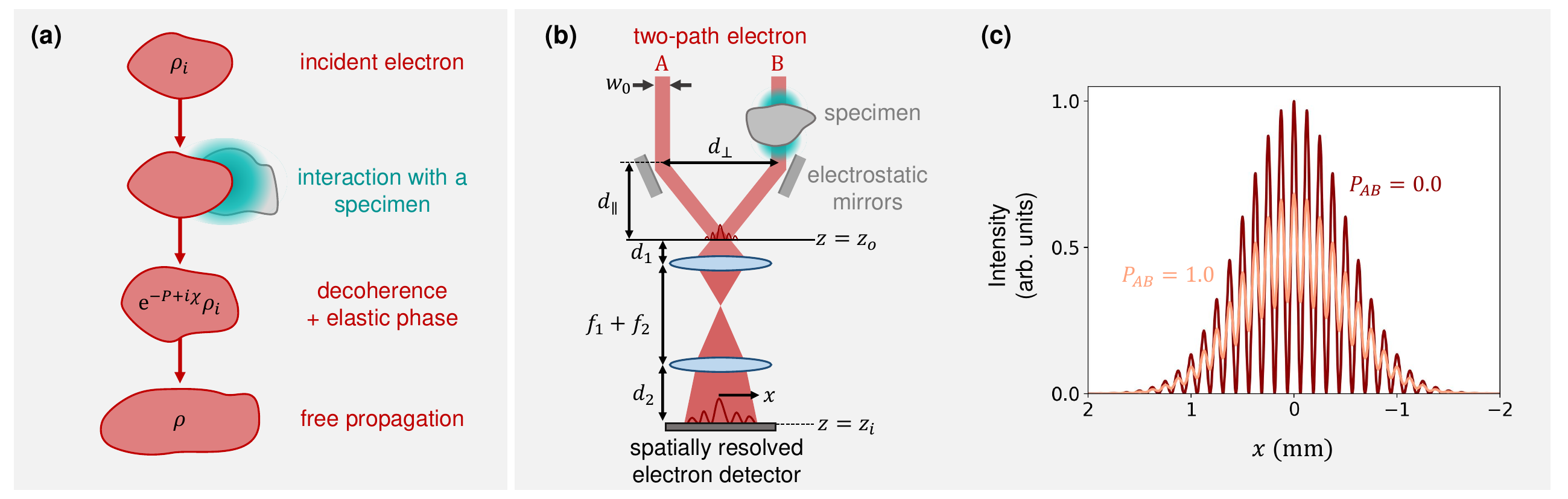}
\caption{\textbf{Free-electron decoherence and interferometry}.
\textbf{(a)}~An incident electron prepared in an arbitrary state corresponding to a density matrix $\rho_i(\rb,\rb')$ interacts with a specimen and evolves according to Eq.~(\ref{evol}), acquiring an elastic phase $\chi$ and undergoing a loss of coherence (probability $P$) associated with inelastic processes.
\textbf{(b)}~We consider an electron prepared in a two-path superposition (interpath distance $d_\perp$, Gaussian path radius $w_0$) such that each path excites the specimen with a different amplitude. Electron paths are then recombined at a plane $z_o$ upon reflection on electrostatic mirrors and propagation over a distance $d_\parallel$. A microscope consisting of two lenses with focal lengths $f_1$ and $f_2$ maps the $z_o$ plane onto an enlarged image at a plane $z_i$, where an electron detector records the magnified interference fringes. The distance $d_1$ from $z_o$ to lens 1 determines the distance $d_2$ from $z_i$ to lens 2 (see Appendix~\ref{apdxprop}), while the magnification factor is fixed by $g=f_2/f_1$.
\textbf{(c)}~Interference patterns formed with 200~keV electrons under the configuration of panel (b) as a function of lateral position $x$ for $g=5000$, $d_\parallel/d_\perp=10^4$, $w_0=200$~nm, and two different values of the interpath decoherence probability $P_{AB}$ with $\chi_{AB}=0$.}
\label{Fig1}
\end{figure*}

Free-electron decoherence can be studied by analyzing the real-space representation of the reduced density matrix $\rho(\rb,\rb')$, obtained from the entire joint electron--environment quantum state by tracing out all degrees of freedom associated with the environment (including radiation and material excitations), assumed to be in thermal equilibrium at a temperature $T$. Starting with an electron moving along the $z$ axis and entering the interaction region in a state $\rho_i(\rb,\rb')$ (see schematic in Fig.~\ref{Fig1}a), the post-interaction density matrix becomes (see Sec.~\ref{sectheory} below and Refs. \cite{paper357,paper425})
\begin{align} \label{evol} 
\rho(\rb,\rb')=\ee^{-P(\rb,\rb')+\ii \chi(\rb,\rb')}\rho_i(\rb,\rb').
\end{align}
Here, $P(\rb,\rb')$ acts as a decoherence probability accounting for an exponential loss of coherence between the transverse spatial components of $\rho_i$, encapsulating all the inelastic interactions undergone by the electron. Also, $\chi(\rb,\rb')$ embodies an elastic phase shift associated with polarization of the environment \cite{paper357}.

In this work, we focus on $P(\rb,\rb')$ and provide a detailed exploration of electron coupling to planar structures made of different materials. We analyze the inelastic processes that lead to decoherence, highlighting the distinction between bulk, surface, and radiative excitations. The article is organized as follows: In Sec.~\ref{sectheory}, we present a general theoretical framework to compute the temperature-dependent decoherence for any arbitrary material and geometry, described in terms of the electromagnetic Green tensor. Additionally, we propose a simple experimental configuration as an example to measure both $P$ and $\chi$. In Secs.~\ref{secbulk} and \ref{secint}, we analyze different sources of decoherence, and in particular, in Sec.~\ref{secbulk}, we study bulk-induced decoherence for an e-beam split into different paths propagating inside an infinitely extended medium made of either Au, Al, or LiF. In Sec.~\ref{secint}, we examine surface effects associated with configurations in which two paths travel either parallel or perpendicular to planar surfaces. In Sec.~\ref{sectemp}, we study decoherence as a function of temperature in the specimen, exploring the temperature sensitivity achieved by measuring the fringe visibility in energy-filtered holography. We conclude in Sec.~\ref{conclusions} by summarizing our findings and discussing the potential of decoherence as a novel probe for thermal sensing, compared with other existing techniques.

\section{Free-electron spatial decoherence}
\label{sectheory}

\subsection{Effect of decoherence on electron interference} 

To study how decoherence influences the evolution of free electrons after the interaction with material structures and scattered electromagnetic fields, we first discuss a simple tutorial scenario that relates closely to electron microscope setups. We consider the configuration sketched in Fig.~\ref{Fig1}b, starting with an e-beam of relativistic electrons prepared in a two-path superposition under paraxial propagation conditions. We assume loosely focused Gaussian sub-beams, such that their transverse profile is preserved without diffraction during propagation along the entire length of the setup (e.g., over the Rayleigh range $z_R=\pi w_0^2/\lambda_e\approx50$~mm for 200~keV electrons and a Gaussian radius $w_0=200$~nm). In addition, we assume a small path width compared to the characteristic spatial variation of elastic and inelastic interactions. One of the paths passes through or close to a specimen, where it acquires some decoherence relative to the other path. In general, interpath decoherence arises when there is a different degree of interaction between each path and the specimen. After interaction, the paths are reflected by electrostatic mirrors with a small deflection angle that preserves paraxial propagation. The two paths overlap after a propagation distance $d_\parallel$, where they display interference fringes. Finally, the overlapping region ({\it object} plane $z_o$) is magnified by a factor $f_2/f_1$ upon transmission through a microscope consisting of two lenses of focal lengths $f_1$ and $f_2$, separated by a distance $f_1+f_2$ (see Fig.~\ref{Fig1}b). An electron detector is placed at the {\it image} plane $z_i$.

For simplicity, we consider an interpath separation $d_\perp\gg w_0$. As anticipated in Eq.~(\ref{evol}), the two electron paths, which we denote $A$ and $B$, acquire a relative decoherence probability $P_{AB}=P_{BA}$, which reduces the visibility of the interference pattern, accompanied by an elastic phase shift $\chi_{AB}=-\chi_{BA}$. The latter produces a lateral shift of the fringes. Furthermore, we neglect intrapath decoherence and phase shifting, assuming that the path radius $w_0$ is small compared with the spatial frequencies of the electromagnetic fields mediating the electron interaction with the specimen. Under these conditions, the intensity recorded in the image plane as a function of lateral position $x$ reduces to (see detailed derivation in Appendix~\ref{apdxprop})
\begin{align} \label{intdemo} 
I(x)\propto \ee^{-2x^2/w_i^2} \Big[ 1+\ee^{-P_{AB}}\cos\big(x/d_i + \chi_{AB}\big)\Big],
\end{align}
where the distances $w_i=gw_0$ and $d_i=(g/2\pi)\,(d_\parallel/d_\perp)\,\lambda_e$ are magnified by the lens factor $g=f_2/f_1$ and the inverse numerical aperture $d_\parallel/d_\perp$.

The contrast of the interference pattern can be quantitatively characterized using the fringe visibility, defined as $V=(I_{\rm max}-I_{\rm min})/(I_{\rm max}+I_{\rm min})$, which measures the normalized difference between the intensities at adjacent maxima and minima. From Eq.~(\ref{intdemo}), we find $V=\ee^{-P_{AB}}$ at the center of the interference pattern (assuming $w_i\gg d_i$). Consequently, increasing the decoherence probability $P_{AB}$ leads to an exponential suppression of the contrast of the interference pattern. While this loss of contrast is usually interpreted as a degradation of the information accessible through interferometric measurements, it can also be exploited as a probe of environmental properties. In particular, since $P_{AB}$ depends on the sample temperature $T$, measurements of the visibility can provide a means of inferring local thermal conditions, as discussed in Sec.~\ref{sectemp}.

\subsection{General theory of electron decoherence} 
\label{generaltheory}

To obtain a general explicit form of Eq.~(\ref{evol}), we represent the electron state through the real-space density matrix $\rho(\rb,\rb')$ and study its evolution during the interaction with materials and scattered electromagnetic fields. We work under the nonrecoil approximation, assuming that the initial momentum uncertainty and the momentum exchanges associated with the interaction are both negligible compared to the central electron momentum $\hbar\qb_0$. The latter is taken along the $z$ direction without loss of generality. We define the central electron kinetic energy $E_0=\me c^2\big[\sqrt{1+(\hbar q_0/\me c)^2}-1\big]$ and velocity $v=c\sqrt{1-1/(E_0/\me c^2+1)^2}$, as well as the Lorentz factor $\gamma=1/\sqrt{1-v^2/c^2}$.

We consider the system formed by the electron (el) and its environment (env). The latter consists of material structures and radiation fields, including polaritons and other excitations. The total Hamiltonian of the system can be written in real-space electron coordinates as $\Hh(\rb) = \Hh_{\rm el}(\rb)+\Hh_{\rm env} +\Hint(\rb)$, which comprises the noninteracting electron and environment components $\Hh_{\rm el}(\rb)$ and $\Hh_{\rm env}$,  as well as the interaction term $\Hint(\rb)$. By linearizing the electron energy $E_\qb \approx E_0 + \hbar \vb\cdot (\qb-\qb_0)$ as a function of wavevector $\qb$ relative to $\qb_0$, we can approximate the electron Hamiltonian as $\Hh_{\rm el}(\rb)=E_0-\hbar \vb\cdot(\qb_0+\ii \nabla)$. The environment consists of electromagnetic modes $i$ of energies $\hbar\omega_i$, which include their interaction with the materials. We write the corresponding Hamiltonian as $\Hh_{\rm env}=\hbar\sum_i\omega_i\ahd_i\ah_i$, where $\ahd_i$ and $\ah_i$ are creation and annihilation operators. We assume that all modes under consideration have a bosonic nature (i.e., $[\ah_i,\ahd_{i'}]=\delta_{ii'}$). Because the electron velocity is constant in the nonrecoil approximation, the spatially integrated electron current reduces to $e\vb$, and therefore, adopting the minimal-coupling prescription, orking in the temporal gauge (scalar potential set to zero), and neglecting $A^2$ terms, the interaction Hamiltonian becomes $\Hint(\rb) = (e\vb/c)\cdot \hat{\Ab}(\rb)$, where $\vb=v\zz$, while $\hat{\Ab}(\rb)=-\ii c\sum_i\omega_i^{-1}\big[\vE_i(\rb)\ah_i-\vE^*_i(\rb)\ahd_i\big]$ is the vector potential operator constructed from the mode electric-field functions $\vE_i(\rb)$. We remark that the latter incorporates the interaction with material structures.

We switch to the interaction picture, where $\hat{\Ab}(\rb)$ becomes time-dependent due to both the unitary transformation $\ee^{\ii \Hh_{\rm env} t/\hbar} \ah_i \ee^{-\ii \Hh_{\rm env} t/\hbar} = \ee^{-\ii \omega_i t}\ah_i$ and the spatial displacement induced by the $-\ii\hbar\vb \cdot \nabla$ term in $\Hh_{\rm el}(\rb)$. In this picture, the interaction Hamiltonian becomes $\Hint(\rb,t) = (e\vb/c)\cdot\hat{\Ab}(\rb+\vb t,t)$, while the evolution of the electron--environment system is governed by the evolution operator $\hat{\mathcal{S}}(\rb,t)$, relating the state at time $t$ to the initial state according to $\ket{\psi(\rb,t)} = \hat{\mathcal{S}}(\rb,t)\ket{\psi(\rb,-\infty)}$. Assuming that the electron and environment are initially uncorrelated, the initial state of the system is described by the product of the density matrix of the incident electron $\rho_i(\rb,\rb')$ and a thermal environment state at a temperature $T$. After the interaction ($t\to \infty$), we trace out the environment to obtain the reduced final density matrix of the electron 
\begin{align} \label{evolgen} 
\rho_{f}(\rb,\rb')=\langle\hat{\mathcal{S}}(\rb,\infty)\hat{\mathcal{S}}^{\dagger}(\rb',\infty)\rangle_T\;\rho_i(\rb,\rb'),
\end{align}
where $\langle{\dots}\rangle_T$ denotes the thermal average over the environmental degrees of freedom. Finally, we recast Eq.~(\ref{evolgen}) into the exponential form of Eq.~(\ref{evol}) by identifying $-P(\rb,\rb')$ and $\chi(\rb,\rb')$ as the real and imaginary parts of $\log\langle\hat{\mathcal{S}}(\rb,\infty)\hat{\mathcal{S}}^{\dagger}(\rb',\infty)\rangle_T$, respectively. 

Using the Magnus expansion \cite{M1954}, the evolution operature can be written as
\begin{align} \label{Smatrix} 
\hat{\mathcal{S}}(\rb,\infty) =& \,\exp\bigg\{\! -\frac{\ii}{\hbar} \int_{-\infty}^{\infty} dt\, \Hint(\rb,t) \\
&-\frac{1}{2\hbar^2} \int_{-\infty}^{\infty}\!\! dt \int_{-\infty}^{t}\!\! dt' \Big[\Hint(\rb,t) ,\Hint(\rb,t')\Big]\bigg\}.
\nonumber
\end{align}
It is important to stress that this expansion only produces contributions up to second order in the interaction Hamiltonian. Indeed, Eq.~(\ref{Smatrix}) is exact because the vector potential operator is linear in the ladder operators, and therefore, the commutator $[\Ah_z(\rb,t),\Ah_z(\rb',t')]$ is a (purely imaginary) c-number, causing higher-order terms in the expansion to vanish. 

Using the Baker--Campbell--Hausdorff formula, the product of the two evolution operators in Eq.~(\ref{evolgen}) can be combined into a single exponential $\ee^{\hat{C}(\rb,\rb')+\ii \chi(\rb,\rb')}$, where $\hat{C}$ contains linear terms in the creation and annihilation operators, whereas $\chi$ collects the scalar phases arising from the commutators. We evaluate the thermal average using the identity $\langle\exp\{\sum_i (c_i^{*} \ahd_i-c_i \ah_i)\}\rangle_T = \exp\{(1/2)\sum_i\langle (c_i^* \ahd_i - c_i\ah_i)^2\rangle_T\}$, which transforms Eqs.~(\ref{evolgen}) and (\ref{Smatrix}) into Eq.~(\ref{evol}) with a decoherence probability and elastic phase given by
\begin{widetext}
\begin{subequations} \label{pandchi} 
\begin{align} \label{pchi} 
&P(\rb,\rb')= \frac{1}{2}\Big( \frac{ev}{\hbar c} \Big)^2 \bigg\langle\Big\{\int_{-\infty}^{\infty} dt\, \big[\Ah_z(\rb-\vb t,t)-\Ah_z(\rb'-\vb t,t)\big]\Big\}^2\bigg\rangle_T, \\
&\chi(\rb,\rb')= \frac{1}{2}\Big( \frac{ev}{\hbar c} \Big)^2 \Big(
\int_{-\infty}^{\infty} dt \int_{-\infty}^{\infty} dt'\, \big[\Ah_z(\rb-\vb t,t),\Ah_z(\rb'-\vb t',t')\big] \\
&\quad\quad\quad\quad\quad\quad\quad
+\int_{-\infty}^{\infty} dt \int_{-\infty}^{t} dt'\,\Big\{\big[\Ah_z(\rb-\vb t,t),\Ah_z(\rb-\vb t',t')\big]-\big[\Ah_z(\rb'-\vb t,t),\Ah_z(\rb'-\vb t',t')\big]\Big\}\bigg). \nonumber
\end{align}
\end{subequations}
\end{widetext}
Both $P$ and $\chi$ are contributed by $A^2$ terms, and we recall that higher-order terms vanish exactly for the bosonic exciations under consideration. Incidentally, the ponderomotive term that we have neglected in the interaction Hamiltonian could also produce an $A^2$ contribution through the linear term in Eq.~(\ref{Smatrix}), but with a relative weight $\sim\lambda_e/L\ll1$ compared to Eqs.~(\ref{pandchi}), where $L$ is the characteristic interaction length determined by both geometry and the spatial extension of the excitations.

The thermal average in Eqs.~(\ref{pandchi}) can be related to the macroscopic electromagnetic properties of the system by applying the fluctuation--dissipation theorem \cite{paper425}
\begin{align} \nonumber 
\big\langle A_z(\rb,t) &A_{z}(\rb',t')\big\rangle_T=-4\hbar c^2 \int_0^{\infty} d\omega\, \Imm\big\{G_{zz}(\rb,\rb',\omega)\big\} \\
&\times \Big\{ 2n_T(\omega) \cos\big[\omega(t-t')\big]+\ee^{-\ii\omega (t-t')} \Big\},
\label{theravg}
\end{align}
where $n_T(\omega)=\big(\ee^{\hbar\omega/\kB T}-1\big)^{-1}$ is the Bose--Einstein distribution at a temperature $T$ and $G_{zz}(\rb,\rb',\omega)$ is the $zz$ component of the $3\times3$ electromagnetic Green tensor. The latter is implicitly defined by
\begin{align} \nonumber 
\nabla\times&\nabla\times G(\rb,\rb',\omega)-k^2 \int d^3\rb''\, \epsilon(\rb,\rb'',\omega) \cdot G(\rb'',\rb',\omega) \\
&=-\frac{1}{c^2} \delta(\rb-\rb') \label{green}
\end{align}
with $k=\omega/c$. Here, the response of the materials enters through the (generally nonlocal) permittivity tensor $\epsilon(\rb,\rb',\omega)$. Note that, in Eq.~(\ref{theravg}), we have taken the continuum limit for the energies of the electromagnetic modes, transforming the discrete sum over $i$ into an integral over $\omega$ [see Ref. \cite{paper425} for a self-contained derivation of Eqs.~(\ref{pandchi}) and (\ref{theravg})]. Assuming we are dealing with reciprocal materials, such that $G_{zz}(\rb,\rb',\omega)=G_{zz}(\rb',\rb,\omega)$, we can use Eq.~(\ref{theravg}) to readily obtain the commutator 
\begin{align} \nonumber 
\big[A_z(\rb,t),A_{z}(\rb',t')\big] = & 8\ii\hbar c^2 \int_0^{\infty} d\omega\, \sin\big[\omega(t-t')\big] \\
&\times \Imm\big\{G_{zz}(\rb,\rb',\omega)\big\}. \label{theravgbis}
\end{align}
In Eq.~(\ref{theravgbis}), we have omitted the thermal average on the left-hand side because the commutator is a c-number. This is corroborated by the fact that the temperature-dependent term in Eq.~(\ref{theravg}) cancels out when subtracting the two terms of the commutator.

Direct application of Eqs.~(\ref{theravg}) and (\ref{theravgbis}) allows us to rewrite Eqs.~(\ref{pandchi}) in a more convenient form in terms of the electromagnetic Green tensor:
\begin{widetext}
\begin{subequations} \label{pandchibis} 
\begin{align}\label{decoP}  
&P(\rb,\rb')= \int_0^\infty d\omega
\;\Big[n_T(\omega)+\frac{1}{2}\Big]
\;\big[\Gamma(\rb,\rb,\omega) +\Gamma(\rb',\rb',\omega) -2\,\Gamma(\rb,\rb',\omega) \big], \\
&\chi(\rb,\rb')=\frac{2e^2}{\hbar} \int_{-\infty}^\infty dz''\int_{-\infty}^\infty dz'''\int_0^\infty d\omega
\;\bigg\{2\sin\Big[\frac{\omega}{v}(z-z'-z''+z''')\Big]
\;\Imm\big\{G_{zz}(\Rb,z'',\Rb',z''',\omega)\big\} \\
&\quad\quad\quad\quad\quad\quad+\cos\Big[\frac{\omega}{v}(z''-z''')\Big]
\;\Ree\big\{G_{zz}(\Rb',z'',\Rb',z''',\omega)-G_{zz}(\Rb,z'',\Rb,z''',\omega)\big\} \bigg\},\nonumber
\end{align}
\end{subequations}
where
\begin{align} \label{geneels} 
\Gamma(\rb,\rb',\omega) = \frac{4e^2}{\hbar}\int_{-\infty}^{\infty} dz'' \int_{-\infty}^{\infty} dz''' \cos\Big[ \frac{\omega}{v} (z-z'-z''+z''')\Big]\, \Imm\big\{-G_{zz}(\Rb,z'',\Rb',z''',\omega)\big\} 
\end{align}
\end{widetext}
acts as a nonlocal electron energy-loss probability. Here, $\Rb=(x,y)$ is the transverse component of $\rb$ relative to the e-beam.

From Eqs.~(\ref{pandchibis}), the decoherence probability is found to be a symmetric function of spatial positions [$P(\rb,\rb')=P(\rb',\rb)$], while the phase is antisymmetric [$\chi(\rb,\rb')=-\chi(\rb',\rb)$]. For $\rb=\rb'$, we have $\chi(\rb,\rb)=0$, while $\Gamma(\rb,\rb,\omega)$ coincides with the loss probability in electron energy-loss spectroscopy (EELS) \cite{paper149}. In addition, we also have $P(\rb,\rb)=0$, which implies that the interaction does not modify the diagonal of the density matrix.

In the nonrecoil approximation here employed, $\Gamma(\rb,\rb',\omega)$ [and, consequently, also $P(\rb,\rb')$] is independent of the longitudinal electron wavefunction (along the e-beam direction) \cite{paper371}. Therefore, for simplicity, we consider a point-like electron (extremely compressed longitudinal wavefunction), which can be treated as a classical current ${\bf j}(\rb,\omega)=-e\,\zz\,\ee^{\ii \omega z/v}\delta(\Rb-\Rb')$, where $\Rb'$ denotes the transverse electron coodinates. In addition, we evaluate $P(\rb,\rb',\omega)=P(\Rb,z,\Rb',z',\omega)$ at the same plane $z'=z$ for both paths, so we drop the dependence on $z$ for clarity. The field induced  at a position $\rb$ by the current above is given by
\begin{subequations}
\label{practicalcalculation}
\begin{align} \nonumber 
\Eb(\rb,\Rb',\omega) = 4\pi\ii e\omega \int_{-\infty}^{\infty} dz'\,\ee^{\ii\omega z'/v}\, G(\rb,\rb',\omega)\cdot \zz,
\end{align}
from which we obtain the decoherence probability by recasting Eq.~(\ref{geneels}) into
\begin{align} \label{eelsfield1} 
\Gamma(\Rb,\Rb',\omega) = \frac{1}{2}\big[\tilde{\Gamma}(\Rb,\Rb',\omega)+\tilde{\Gamma}(\Rb',\Rb,\omega)\big]
\end{align}
with
\begin{align} \label{gammafield} 
\tilde{\Gamma}(\Rb,\Rb',\omega) = \frac{e}{\pi\hbar\omega}\int_{-\infty}^{\infty} dz\; \Ree\big\{ \ee^{-\ii \omega z/v} E_{z}(\rb,\Rb',\omega)\big\}.
\end{align}
\end{subequations}
This result provides a clearer physical interpretation of $\Gamma(\Rb,\Rb',\omega)$, which represents the probability that two electrons at lateral positions $\Rb$ and $\Rb'$ exchange quanta of energy $\hbar\omega$. For clarity, we stress that we are considering single electrons throughout this work (i.e., low e-beam currents, as employed in electron microscopes), so $\Gamma(\Rb,\Rb',\omega)$ refers to different lateral positions of the same electron. We note that Eqs.~(\ref{practicalcalculation}) are suitable for numerical implementation (e.g., using boundary-element methods \cite{paper040}), enabling decoherence modeling in geometries where analytical approaches are unavailable.

The EELS and decoherence probabilities can exceed unity, but this does not represent a physical problem. In particular, the decoherence probability enters through an exponential in Eq.~(\ref{evol}), and thus, large values exceeding unity must be understood as a substantial depletion of coherence. Likewise, the EELS probability must be understood as an average number of excitations produced per electron. For example, considering a single bosonic mode initially prepared in the ground state, the post-interaction populations of number states follow a Poissonian distribution whose average is given by the frequency-integrated EELS probability \cite{LKB1970,SL1971,paper228}.

For our two-path electron (Fig.~\ref{Fig1}b), decoherence becomes significant when the excitations generated by path $A$ (lateral position $\Rb_A$) are sufficiently different from those produced by path $B$ (at $\Rb_B$). The decoherence probability [Eq.~(\ref{decoP})] grows with the local EELS probabilities at $\Rb_A$ and $\Rb_B$ ($\Rb=\Rb'$ terms), while the cross-term ($\Rb\neq\Rb'$) mitigates decoherence for modes excited with similar amplitude and phase at both positions. Strongly localized modes, specifically those with wavelengths or propagation lengths much shorter than $|\Rb-\Rb'|$, are therefore expected to dominate the decoherence process. This motivates a spectral decomposition of the decoherence probability. Consequently, we define the frequency-resolved decoherence probability $P(\Rb,\Rb',\omega)$ as the integrand in Eq.~(\ref{decoP}), such that $P(\Rb,\Rb') = \int_0^{\infty}d\omega\, P(\Rb,\Rb',\omega)$.

Besides the Bose--Einstein factor in Eq.~(\ref{decoP}), there is a temperature dependence in the dielectric properties of the materials that form the specimen through the combined effect of lattice expansion, reduced excitation lifetimes, and changes in the electronic band structures. Although we do not enter into the details of these additional effects, we comment briefly on their relative importance. In metals, thermal lattice expansion and changes in the effective electron mass induce a small shift in the plasma frequency, while enhanced electron--phonon scattering significantly increases the damping rate \cite{IN1970,RGK16}. In dielectrics such as LiF, where lattice vibrations dominate the response, thermal expansion weakens the effective lattice force constants, causing a redshift in the resonance frequencies. Simultaneously, temperature-enhanced anharmonic phonon--phonon scattering increases optical damping \cite{G1960,JKP1966}. Finally, in transition metal dichalcogenides (e.g., MoS$_2$), excitons dominate the visible optical response, and temperature changes produce substantial redshifts and broadenings of the resonance peaks via bandgap renormalization and enhanced exciton--phonon scattering \cite{TZA12,HCK15,SBR16}. Additionally, thermal energy can accelerate exciton dissociation into free carriers, resulting in a loss of spectral weight \cite{TZA12}. All of these thermal effects can be readily incorporated into our formalism by employing temperature-dependent permittivities.

\section{Electron decoherence in bulk media}
\label{secbulk}

We study the decoherence experienced by an electron propagating inside an infinite homogeneous medium characterized by a momentum- and frequency-dependent dielectric tensor $\epsilon(\qb,\omega)$. We separate the wavevector $\qb=\Qb+q_z\zz$ into transverse and longitudinal components $\Qb=(q_x,q_y)$ and $q_z$ relative to the propagation direction $z$. Assuming translational invariance, the dielectric tensor can be written
\begin{align} \label{epstens} 
\epsilon(\qb,\omega) = &\epsilon_{\rm lon}(\qb,\omega)\; \qq\otimes\qq \\
&+\epsilon_{\rm tr}(\qb,\omega)\; (\eb_{\Qb s} \otimes \eb_{\Qb s} + \eb_{\Qb p} \otimes \eb_{\Qb p}) 
\nonumber
\end{align}
in terms of longitudinal (lon) and transverse (tr) components. Here, $\hat{\qb}=\qb/q$, $\eb_{\Qb s}=\zz\times\QQ=(-q_y\xx+q_x\yy)/Q$, $\eb_{\Qb p}=\eb_{\Qb s}\times\qq=(q_z\QQ-Q\zz)/q$, and $\QQ=\Qb/Q$. The corresponding real-space representation is $\epsilon(\rb,\rb',\omega) = (2\pi)^{-3}\int d^3 \qb\, \ee^{\ii \qb\cdot(\rb-\rb')}\, \epsilon(\qb,\omega)$ [see Eq.~(\ref{green})].

\begin{figure*}
\centering\includegraphics[width=1.0\textwidth]{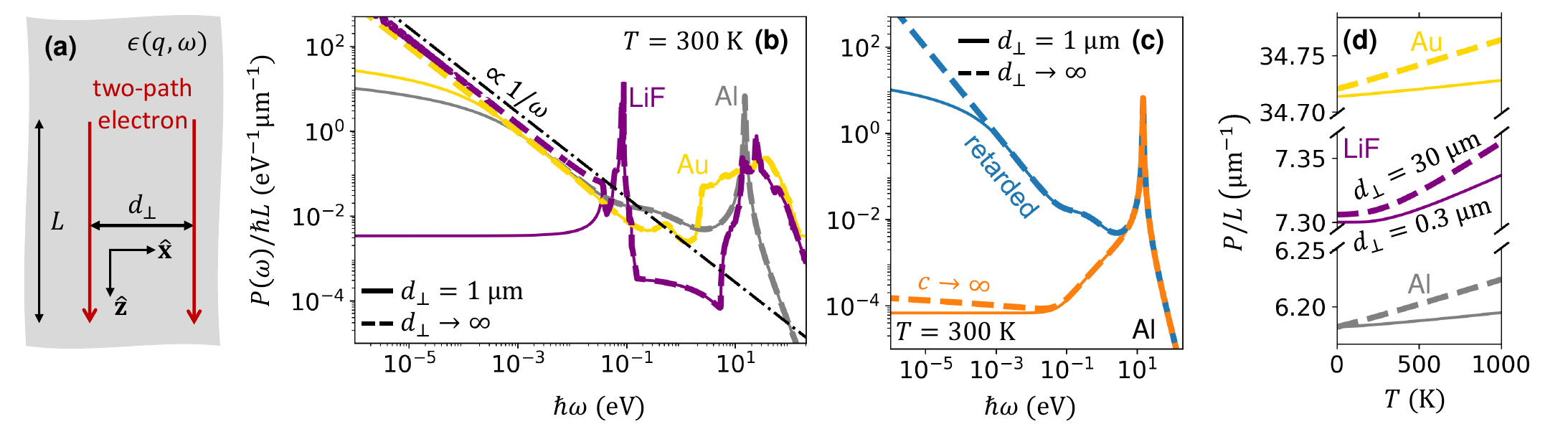}
\caption{\textbf{Bulk contribution to electron decoherence}.
\textbf{(a)}~An electron prepared in a two-path superposition propagates along a distance $L$ inside the bulk of a material characterized by a nonlocal dielectric tensor $\epsilon(q,\omega)$. The two paths are separated by a distance $d_\perp$.
\textbf{(b)}~Frequency-resolved decoherence probability $P(\omega)$ for a $200$~keV electron under the configuration of panel (a) when the material is Au, Al, or LiF at room temperature ($T=300$~K). We consider $d_\perp=1$~$\mu$m (solid curves) and $d_\perp\to\infty$ (dotted curves, coinciding with the EELS probability for a single-path electron).
\textbf{(c)}~Comparison between the decoherence probability obtained for Al with inclusion of retardation [taken from (b)] and the nonretarded calculation (i.e., taking $c\to\infty$).
\textbf{(d)}~Temperature dependence of the frequency-integrated decoherence probability $P$ for $d_\perp=0.3$~$\mu$m (solid curves) and $d_\perp=30$~$\mu$m (dashed curves). The half-collection angle is set to $\varphi_{\rm out}=10$~mrad throughout this work.}
\label{Fig2}
\end{figure*}

We evaluate the decoherence probability by applying Eqs.~(\ref{decoP}) and (\ref{geneels}) to an electron prepared in a coherent superposition of two straight-line paths separated by a distance $d_\perp$ (Fig.~\ref{Fig2}a). To this end, we first obtain the Green tensor $G=G^{\rm bulk}$ by rewriting Eq.~(\ref{green}) in momentum space as $\qb\times\qb\times G^{\rm bulk}(\qb,\omega)+k^2 \epsilon(\qb,\omega) G^{\rm bulk}(\qb,\omega)=1/c^2$. For the dielectric tensor in Eq.~(\ref{epstens}), we find the solution
\begin{align} 
G^{\rm bulk}(\qb,\omega) = \frac{\hat{\qb} \otimes \hat{\qb}}{\omega^2\epsilon_{\rm lon}(\qb,\omega)} + \frac{\hat{\bf{e}}_{\Qb s}\otimes\hat{\bf{e}}_{\Qb s} + \hat{\bf{e}}_{\Qb p}\otimes\hat{\bf{e}}_{\Qb p}}{[\omega^2\epsilon_{\rm tr}(\qb,\omega)-c^2q^2]}.
\nonumber
\end{align}
We now proceed by moving back to real space via $G^{\rm bulk}(\rb,\rb',\omega)=(2\pi)^{-3}\int d^3 \qb\,\ee^{\ii\qb\cdot(\rb-\rb')}\, G^{\rm bulk}(\qb,\omega)$, retaining the $zz$ component, inserting it into Eq.~(\ref{geneels}) to obtain the nonlocal loss-probability, and substituting the result into Eq.~(\ref{decoP}). We then obtain the {\it bulk} decoherence probability
\begin{align} \label{Pbulk} 
&P^{\rm bulk}(d_\perp) \\
&= \int_0^{\infty} d\omega\,[2n_T(\omega)+1]\, [\Gamma^{\rm bulk}(0,\omega)-\Gamma^{\rm bulk}(d_\perp,\omega)],
\nonumber
\end{align}
where
\begin{widetext}
\begin{align}
\Gamma^{\rm bulk}(R,\omega)=&\frac{2e^2L}{\pi\hbar v^2} \int_0^{\infty} Q\, dQ J_0(QR) \; \Imm\Bigg\{\Bigg[\frac{1}{\epsilon_{\rm tr}(q,\omega)}-\frac{1}{\epsilon_{\rm lon}(q,\omega)}\Bigg]\frac{1}{q^2}+\Bigg[\frac{v^2}{c^2}-\frac{1}{\epsilon_{\rm tr}(q,\omega)}\Bigg]\frac{1}{q^2-k^2\epsilon_{\rm tr}(q,\omega)}\Bigg\}
\label{gammabulk}
\end{align}
\end{widetext}
is the nonlocal energy-loss probability for an electron moving in a bulk medium and two positions separated by a distance $R$ \cite{paper149}.

To study a representative set of materials, we apply Eqs.~(\ref{Pbulk}) and (\ref{gammabulk}) to a dielectric (LiF), as well as simple (Al) and noble (Au) metals, where we explore the contributions of phononic and plasmonic modes, respectively. For LiF, we adopt the local approximation and assume isotropic permittivity components $\epsilon_{\rm lon}(\omega) = \epsilon_{\rm tr}(\omega)$. We rely on experimental data from Ref.~\cite{P1985}, which we extend to lower frequencies by fitting the optical response to a two-oscillator model as
\begin{align} \label{epslif} 
\epsilon^{\rm LiF}(\omega) = \epsilon_\infty + \sum_{i = 1}^{2} \frac{s_i \omega_i^2}{\omega_i^2-\omega(\omega+\ii \eta_i)},
\end{align}
with parameters $\epsilon_\infty = 1.96$, $\hbar\omega_1 = 38$~meV, $\hbar\omega_2 = 62$~meV, $\hbar\eta_1 = 2.16$~meV, $\hbar\eta_1 = 10.7$~meV, $s_1 = 6.67$, and $s_2 = 0.116$ \cite{P1985}. We specifically apply Eq.~(\ref{epslif}) to frequencies below the lowest-energy phonon peak at $\hbar\omega = 38$~meV, noting that it reproduces the experimental data accurately near the phonon resonances, although it does not capture higher-frequency features, which become increasingly relevant beyond LiF's band gap of $13.6$~eV.

For metals, nonlocal effects may become relevant due to high-momentum transfers associated with the strong lateral focusing of the electron. To capture these effects, we adopt an independent-electron picture for the conduction electrons, whose longitudinal response is modelled in the random-phase approximation (RPA) \cite{L1954}, corrected to conserve the local electron density as prescribed by Mermin \cite{M1970}. We thus set $\epsilon_{\rm lon}(q,\omega)$ to the Mermin dielectric function $\epsilon^{\rm M}(q,\omega)$ (see Appendix \ref{apdxeps}). The transverse component, which accounts for polarization associated with longer-wavelength modes, is approximated by a local Drude model
\begin{align} \label{drude} 
\epsilon^{\rm D}(\omega)=1-\frac{\wp^2}{\omega(\omega+\ii \eta)},
\end{align}
where $\wp$ is the classical plasma frequency and $\eta$ is the damping rate. We set $\hbar\omega_p=15$~eV and $\hbar\eta=0.6$~eV for Al \cite{P1985}, and $\hbar\omega_p = 9$~eV and $\hbar\eta = 0.05$~eV for Au \cite{JC1972}. 
Note that the Mermin function rigorously reduces to the Drude form in the local limit [i.e., $\epsilon^{\rm M}(q\to0,\omega) =\epsilon^{\rm D}(\omega)$]. This model produces excellent results for Al, where the response of inner bands plays a negligible role.

In contrast to Al, valence-band screening becomes important in Au, particularly in the plasmonic region. To address this issue, we adopt a model that treats the valence-band response in the local approximation and conduction electrons in the Lindhard+Mermin model \cite{paper119}. Specifically, we estimate the contribution of valence bands to the permittivity from the difference between the experimentally measured frequency-dependent dielectric function $\epsilon^{\rm exp}(\omega)$ \cite{JC1972} and its fit to the Drude model. In addition, we retain the Mermin model to describe conduction electrons. The resulting longitudinal dielectric function takes the form \cite{paper119} 
\begin{align} \label{epsau} 
\epsilon^{\rm Au}_{\rm lon}(q,\omega) = \epsilon^{\rm exp}(\omega) - \epsilon^{\rm D}(\omega) + \epsilon^{\rm M}(q,\omega).
\end{align}
In addition, the transverse permittivity of Au is identified as $\epsilon^{\rm Au}_{\rm tr}(q,\omega)=\epsilon^{\rm exp}(\omega)$.

In the local approximation, the integral over $Q$ can be performed analytically for a finite path separation $(R>0)$, yielding the result
\begin{align} \label{bulkan1} 
\Gamma^{\rm bulk}(R,\omega) = \frac{2e^2L}{\pi\hbar v^2} \Imm\bigg\{ \bigg(\frac{v^2}{c^2}-\frac{1}{\epsilon}\bigg) K_0\bigg(\frac{\omega R}{v\gamma_\epsilon}\bigg)\bigg\},
\end{align}
where $\gamma_\epsilon = 1/\sqrt{1-\epsilon(\omega)v^2/c^2}$, $\epsilon$ is understood to have an implicit depedence on $\omega$, $K_0$ is a Bessel function of the second kind, and we have used the identity $\int_0^{\infty} dQ \, Q J_0(Q R)/(Q^2+a^2)=K_0(aR)$ (see Eq.~6.532-4 in Ref. \cite{GR1980}). This expression diverges logarithmically in the $R\to0$ limit due to the $\propto Q^{-1}$ scaling of the integrand for large wavevector $Q$, which can take arbitrarily large values for a point-like electron. In practice, momentum exchanges must remain finite, and we introduce a cutoff $Q_c\approx\varphi_{\rm out}q_0$ defined by the half-collection angle of the electron microscope $\varphi_{\rm out}$ \cite{paper149}, which we set to $10$~mrad from here on. By integrating up to $Q_c$ for $R=0$, we find
\begin{align} \label{bulkan2} 
\Gamma^{\rm bulk}(0,\omega)=&\frac{e^2L}{\pi\hbar v^2} \Imm\bigg\{\bigg(\frac{v^2}{c^2}-\frac{1}{\epsilon}\bigg
)\!\log\bigg[1+\bigg(\frac{v\gamma_\epsilon Q_c}{\omega}\Big)^2\bigg]\bigg\}.
\end{align}
Throughout this paper, we introduce this cutoff to obtain results within the local approximation.

For an infinite path separation ($d_\perp \to \infty$), the decoherence probability exhibits a $\omega^{-1}$ divergence for all three materials, as shown in Fig.~\ref{Fig2}b, where we plot the spectral dependence of $P^{\rm bulk}(d_\perp,\omega)$ for $200$~keV electrons. This can be understood by examining the dielectric functions for both types of materials in the low frequency limit [Eq.~(\ref{epslif}) for LiF and Eq.~(\ref{drude}) for the metals]. For LiF, the static permittivity $\epsilon^{\rm LiF}(0)= \epsilon_{\infty}+s_1+s_2$ is purely real, and therefore, $\Gamma^{\rm bulk}$ vanishes unless we have Cherenkov losses under the condition $v/c>1/\sqrt{\epsilon^{\rm LiF}(0)}\;\approx0.33$, which makes $\gamma_\epsilon$ imaginary. Because we consider $200$~keV electrons ($v/c\approx0.7$) throughout this work, the Cherenkov condition at zero frequency is always satisfied. In contrast, for metals, the permittivity has a divergent imaginary part at low frequencies, and thus, both $K_0$ in Eq.~(\ref{bulkan1}) and the logarithm in Eq.~(\ref{bulkan2}) acquire nonzero imaginary parts, regardless of the electron velocity; a divergence is thus observed at low $\omega$ when multiplying the zero-temperature EELS probability by $n_T(\omega\to0)\propto\omega^{-1}$. Interestingly, this divergence is a retardation effect, as it vanishes in the $c\to \infty$ limit, as shown in Fig.~\ref{Fig2}c, where the low-frequency behavior is dramatically different with and without inclusion of retardation. Although we only compare retarded and nonretarded limits for Al here, Au follows a similar trend. Conversely, for LiF, the nonretarded limit makes the Cherenkov condition impossible to be satisfied, thereby suppressing the divergence.

For finite path separations, the divergence is regularized by the cross term in Eq.~(\ref{decoP}), which suppresses contributions from low-energy excitations when their wavelengths are large compared with $d_\perp$, so that the electron--mode interaction produces similar amplitudes in the two paths (i.e., they do not provide which-way information). This behavior is clearly observed in Fig.~\ref{Fig2}b. Incidentally, $P(\omega\to 0,d_\perp)$ is much larger for metals than for LiF, as expected from the fact that $\Imm\{\epsilon\}$ diverges as $\omega^{-1}$ in the former, while it vanishes in the latter.

The $\omega^{-1}$ divergence in the limit of infinite path separation is also a finite-temperature effect, as it originates from thermal occupation of modes with energies below $\kB T$, and indeed, the divergence disappears in the $T=0$ limit (see Supplementary Fig.~\ref{FigS1}). Consequently, the temperature dependence primarily enters through the low-frequency part of the spectrum.

In addition, $P(\omega)$ displays different material-dependent features: for Al, the only discernible peak is the bulk plasmon at $\sim 15$~eV, while Au displays a broader response for energies starting at its bulk plasmon of $\sim 2.5$~eV; for LiF, we find two peaks corresponding to optical phonons at $\sim50$~meV and interband transitions above the electronic band gap of $13.6$~eV.

\begin{figure}
\centering\includegraphics[width= 0.5\textwidth]{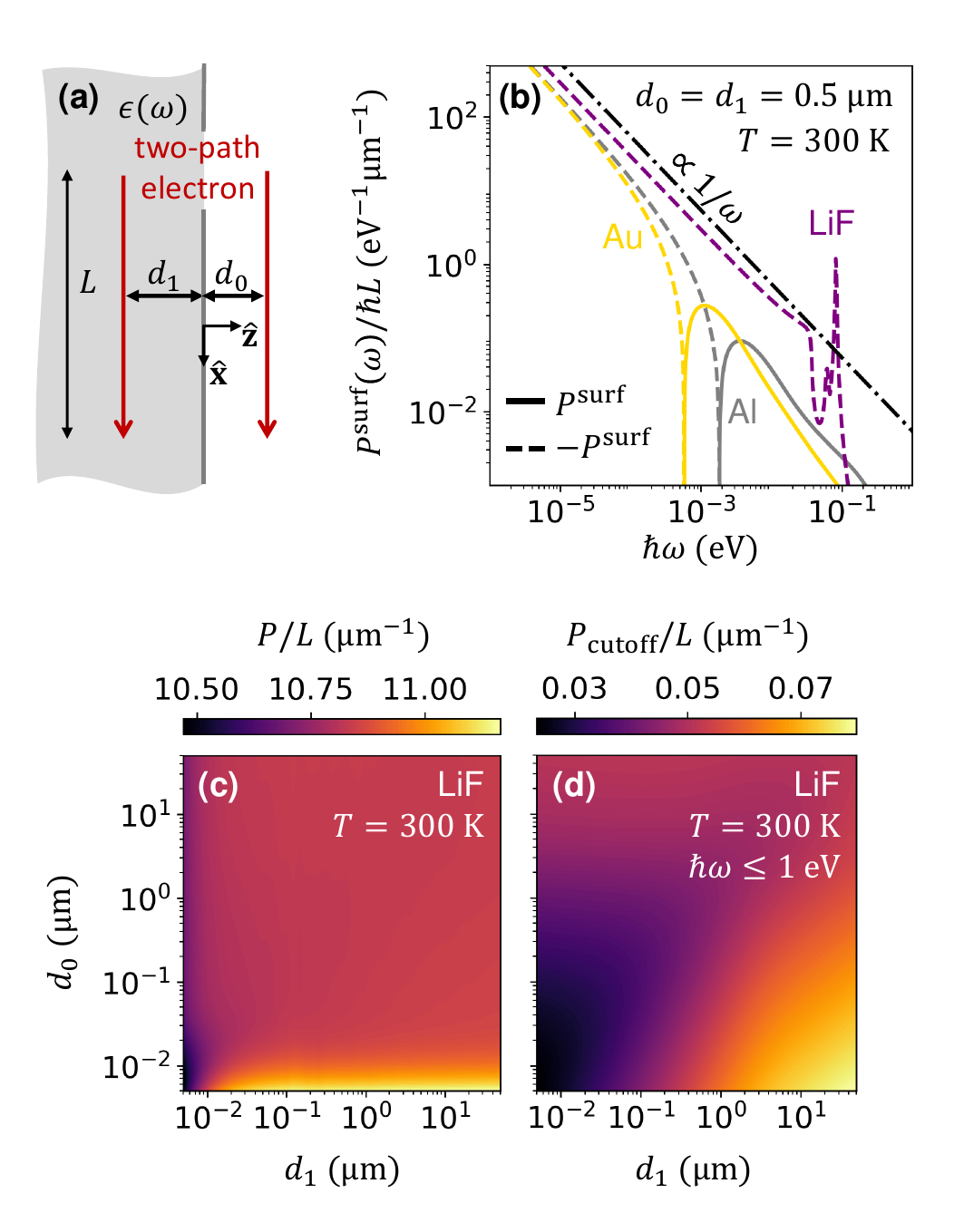}
\caption{\textbf{Surface contribution to electron decoherence: Parallel e-beam configuration}.
\textbf{(a)}~An electron prepared in a two-path superposition propagates along a distance $L$ parallel to the planar surface of a semi-infinite medium of local permittivity $\epsilon(\omega)$. One of the electron paths is inside the material, and the other is in the surrounding vacuum. The path--surface separations are $d_1$ and $d_0$, respectively.
\textbf{(b)}~Surface contribution to the frequency-resolved decoherence probability $P^{{\rm surf}}(\omega)$ for a $200$~keV electron under the configuration of panel (a) when the material is Au, Al, or LiF at room temperature ($T=300$~K). We set $d_0=d_1=0.5~\mu$m. As $P^{{\rm surf}}$ becomes negative for some frequencies, we plot both $P^{\rm surf}$ (solid curves) and $-P^{{\rm surf}}$ (broken curves). The broken black line indicates a $\omega^{-1}$ slope, matching that of the infrared divergence, which is regularized when the bulk contributions are added.
\textbf{(c)}~Decoherence probability $P$ (frequency-integrated bulk+surface) for LiF at $T = 300$~K as a function of $d_0$ and $d_1$.
\textbf{(d)}~Partial decoherence probability $P_{\rm cutoff}$ obtained by limiting the spectral integral to $\hbar\omega\le1$~eV under the same conditions as in panel (c).}
\label{Fig3}
\end{figure}

\section{Surface contribution to electron decoherence}
\label{secint}

The linearity of Eq.~(\ref{green}) allows us to decompose the Green tensor as $G = G^{\rm bulk} + G^{\rm surf}$, where $G^{\rm bulk}$ is defined in each homogeneous region of the specimen as the solution in an infinite medium (e.g., in vacuum or inside a material), while $G^{\rm surf}$ compensates for the presence of surfaces and interfaces. Consequently, since Eq.~(\ref{geneels}) for the nonlocal EELS probability is linear in the Green tensor, we can write it as the sum of bulk and surface contribution $\Gamma=\Gamma^{\rm bulk}+\Gamma^{\rm surf}$, and simnilarly, from Eq.~(\ref{decoP}) for the decoherence probability, $P=P^{\rm bulk}+P^{\rm surf}$. While in Sec.~\ref{secbulk} we have discussed bulk effects for electrons propagating fully inside the bulk of a material, in the present section we study surface effects under two representative configurations for two-path electrons, depending on whether the e-beam is oriented parallel or perpendicular to the surface. In what follows, we include the bulk contribution using the formalism introduced in Sec.~\ref{secbulk}, while surface contribution are calculated within the local approximation, using Eq.~(\ref{epslif}) for the permittivity of LiF, the Drude approximation in Eq.~(\ref{drude}) for Al, and measured data of $\epsilon_{\rm exp}(\omega)$ \cite{JC1972} for Au.

\subsection{Surface-parallel electron beams}
\label{secpar}

We consider the configuration sketched in Fig.~\ref{Fig3}a, where a two-path electron propagates a distance $L$ along $x$ (instead of $z$, which is reserved for the surface normal below), parallel to the surface of a semi-infinite material ($z=0$ plane), with one path moving inside the medium ($z<0$ region) and the other one in vacuum. The electron can couple to surface excitations and also Cherenkov radiation inside the material \cite{paper052}.

Bulk components are directly included in the inner path following  Sec.~\ref{secbulk}, while surface contribution require some analysis of the corresponding electromagnetic Green tensor, which can be obtained by solving Eq.~(\ref{green}) and imposing the the continuity of $x-y$ components of $G(\rb,\rb',\omega)$ and $\nabla\times G(\rb,\rb',\omega)$ at $z=0$ \cite{B12_2}. Alternatively, by noticing that this quantity coincides with the electric field produced at $\rb$ by a dipole of strength $-1/(4\pi\omega^2)$ placed at $\rb'$ and oscillating with frequency $\omega$, we can also obtain the Green tensor from the dipole-induced electric field projected in in-plane momentum components. Following this procedure, we obtain the $xx$ component (i.e., along the e-beam direction) of the surface contribution as
\begin{align} \label{Ginout} 
G^{{\rm surf}}_{xx}(\rb,\rb',\omega) = \frac{-\ii}{8\pi^2 \omega^2}\int \frac{d^2\Qb}{Q^2} \,\ee^{\ii \Qb \cdot (\Rb-\Rb')}\,g(\Qb,z,z',\omega),
\end{align}
where
\begin{widetext}
\begin{align} \label{Ginoutbis} 
g(\Qb,z,z',\omega) = 
\begin{cases} 
    \ee^{\ii q_{z0}(z+z')} (r_s^{01} q_y^2 k^2-q_x^2q_{z0}^2r_{p}^{01})/q_{z0}, & z,z'>0, \\ \\
    \ee^{\ii (q_{z0}z-q_{z1} z')} (t_{s}^{10} q_y^2 k^2 + Q_x^2 q_{z0}q_{z1} t_{p}^{10}/\sqrt{\epsilon})/q_{z1}, & z>0,z'<0, \\ \\
    \ee^{-\ii q_{z1}(z+z')} (r_s^{10} q_y^2 k^2-q_x^2q_{z1}^2r_{p}^{10}/\epsilon)/q_{z1}, & z,z'<0.
\end{cases}
\end{align}
\end{widetext}
Here, $q_{zj}=\sqrt{k^2\epsilon_j-Q^2}$ (with the square root yielding a positive imaginary part) are the out-of-plane wavevector components in vacuum ($j=0$, $\epsilon_0=1$) and inside the material ($j=1$, $\epsilon_1\equiv\epsilon$). Also, we use Fresnel's reflection and transmission coefficients
\begin{align} \nonumber 
&r_s^{jj'}=\frac{q_{zj}-q_{zj'}}{q_{zj}+q_{zj'}}, &r_p^{jj'}=\frac{\epsilon_{j'}\,q_{zj}-\epsilon_jq_{zj'}}{\epsilon_{j'} q_{zj}+\epsilon_jq_{zj'}}, \\
&t_s^{jj'}=\frac{2q_{zj}}{q_{zj}+q_{zj'}}, &t_p^{jj'}=\frac{2\sqrt{\epsilon_j\epsilon_{j'}}\,q_{zj}}{\epsilon_{j'} q_{zj}+\epsilon_jq_{zj'}}, \nonumber
\end{align}
for s and p polarization and incidence from medium $j$ on the $j/j'$ interface. For $z<0$ and $z'>0$, reciprocity allows us to write $G^{{\rm surf}}_{xx}(\rb,\rb',\omega)=G^{{\rm surf}}_{xx}(\rb',\rb,\omega)$.

Inserting Eq.~(\ref{Ginout}) into Eq.~(\ref{geneels}), we obtained the nonlocal EELS probability
\begin{align} \nonumber 
\Gamma^{{\rm surf}}(z,z',\omega) = \frac{2Le^2}{\pi\hbar\omega^2}\int_0^\infty \frac{dq_y}{Q^2} \,
{\rm Re}\big\{g(\omega/v,q_y,z,z',\omega)\big\},
\end{align}
where the condition $q_x=\omega/v$ in $g$ [Eq.~(\ref{Ginoutbis})] is imposed by phase-matching in the electron--surface interaction for a parallel trajectory, producing an overall factor $L$ for the interaction length. Additionally, bulk processes only contribute to the inner path through $\Gamma^{\rm bulk}(z,z,\omega)$ with $z<0$. Finally, the decoherence probability is given by inserting these expressions into Eq.~(\ref{decoP}).

In Fig.~\ref{Fig3}b, we show the frequency-resolved decoherence probability contributed by surface excitations, which can take negative values, although the overall probability including bulk processes must remain non-negative. This effect is particularly relevant at low frequencies, where the surface contribution cancels the divergence of the bulk term $P^{\rm bulk}(\omega)$ for finite path separations. In the infinite separation limit ($|z-z'|=d_\perp\to \infty$), the cross term vanishes in Eq.~(\ref{decoP}), and the spectrally resolved decoherence probability retains an infrared divergence as $\omega^{-1}$ for $T\neq 0$ (see Supplementary Fig.~\ref{FigS3} for results in the zero-temperature limit). Aside from this low-frequency behavior, the results are qualitatively similar for a configuration in which both paths are inside the material, but with a notable offset in the overall value of $P$.

A high decoherence probability under the configuration of Fig.~\ref{Fig3} indicates a higher likelihood of interaction with bulk modes for the path inside the material compared to the one in vacuum. An interesting behavior is observed when comparing excitations from different parts of the spectrum by discarding transmitted electrons that have lost energy above a certain cutoff. In LiF, without any energy cutoff (Fig.~\ref{Fig3}c), $P$ remains nearly constant over a wide range of $d_0$ and $d_1$ values because it is dominated by electronic excitations above the band gap (i.e., $\hbar\omega>13.6$~eV, see Fig.~\ref{Fig2}b), which are strongly localized (i.e., their associated free-space wavelength is $\lambda<90$~nm). In contrast, the decoherence probability $P_{\rm cutoff}$ for a cutoff $\hbar\omega=1$~eV (Fig.~\ref{Fig3}d) exhibits a strong spatial dependence over distances in the micrometer range, as expected from the dominant contributions below $\sim100$~meV, or equivalently $\lambda>12\,\mu$m. With and without a cutoff in $P$, decoherence is maximum when the external path lies near the surface, and the inner one lies deep inside the material, such that the former couples predominantly to surface modes that decay quickly inside the material, while the latter interacts primarily with bulk modes that do not reach the surface.

\begin{figure*}
\centering\includegraphics[width=1.0\linewidth]{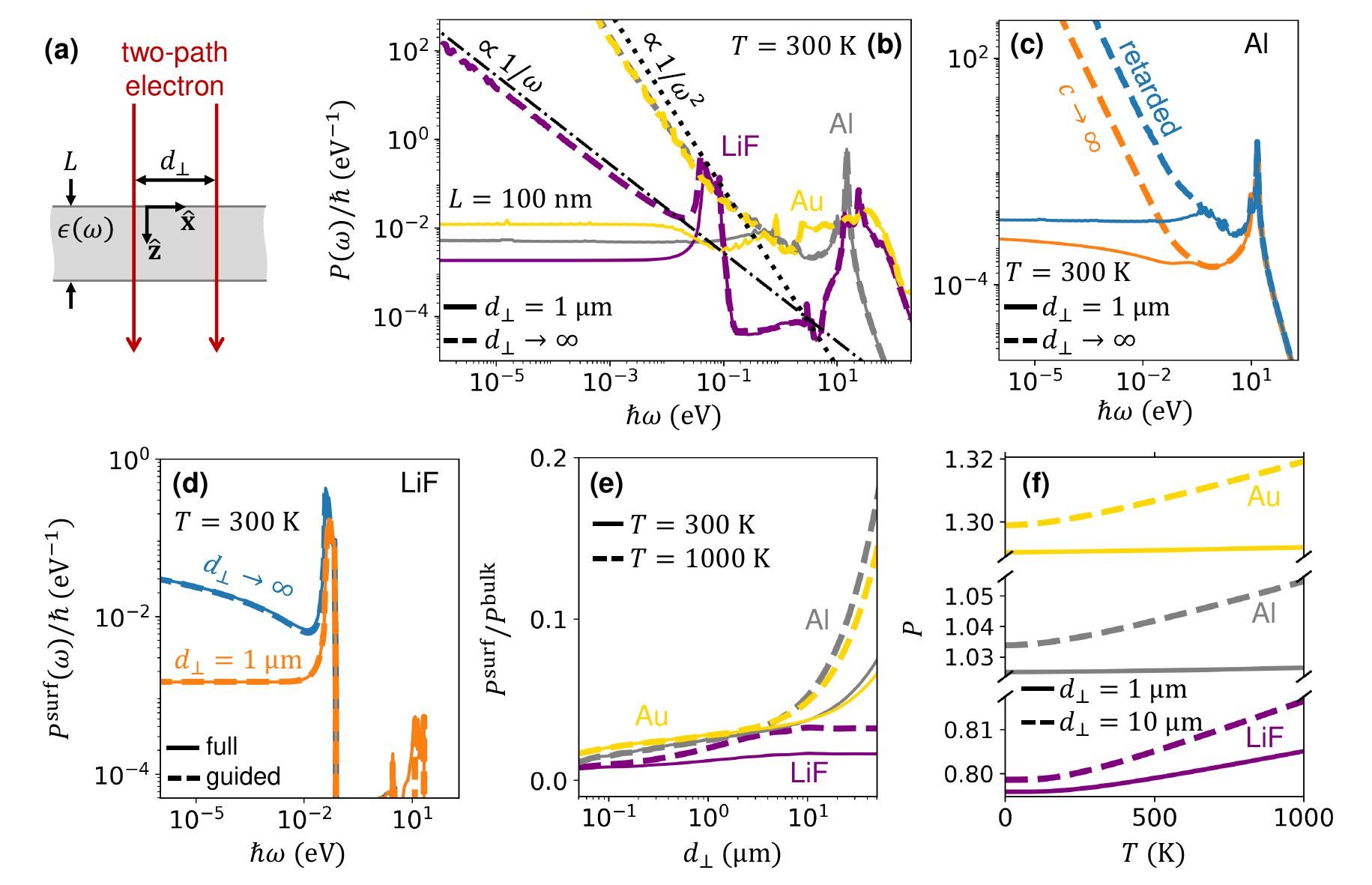}
\caption{\textbf{Surface contribution to electron decoherence: Perpendicular e-beam configuration.}
\textbf{(a)}~An electron prepared in a superposition of two paths separated by a distance $d_\perp$ traverses a homogeneous film (thickness $L$) made of a material of local permittivity $\epsilon(\omega)$. We consider normal e-beam incidence relative to the film.
\textbf{(b)}~Frequency-resolved decoherence probability at $T=300$~K for a $200$~keV electron under the configuration of panel (a) when the material is Au, Al, or LiF. We set $d_\perp=1\,\mu$m (solid curves) and $d_\perp\to\infty$ (broken curves, coinciding with the EELS probability).
\textbf{(c)}~Comparison of retarded and nonretarded regimes for the frequency-resolved decoherence probability in an Al film.
\textbf{(d)}~Full surface contribution to the frequency-resolved decoherence probability $P^{\rm surf}(\omega)$ (solid curves) compared with the partial contribution to this quantity arising from guided modes (dashed curves).
\textbf{(e)}~Ratio $P^{\rm surf}/P^{\rm bulk}$ of the frequency-integrated surface to bulk decoherence probabilities as a function of $d_\perp$ for different values of the temperature.
\textbf{(f)}~Temperature dependence of the decoherence probability for $d_\perp = 10\,{\rm \mu m}$.}
\label{Fig4}
\end{figure*}

\subsection{Thin films traversed under normal incidence}
\label{secperp}

We consider a scenario of particular relevance to holography experiments in transmission electron microscopy, where a two-path electron with transverse separation $d_\perp$ traverses a planar film of thickness $L$ under normal incidence conditions (Fig.~\ref{Fig4}a). The electron interacts with both surface and guided modes supported by the film. In addition, because surface crossing breaks momentum conservation, it also couples to radiative modes, generating transition radiation \cite{GF1946}.

In this geometry, the explicit expression for the Green tensor becomes cumbersome, so instead we compute the decoherence probability using Eq.~(\ref{gammafield}), based on the electron-induced electric field. Given the two-dimensional symmetry of the system, the scattering problem can be solved independently for each in-plane wavevector component $\Qb$, so we expand the field produced by an electron following a trajectory with transverse coordinates $\Rb'$ as $\Eb(\rb,\omega)=(2\pi)^{-2}\int d^2\Qb\,\ee^{\ii\Qb\cdot(\Rb-\Rb')}\,\Eb(\Qb,z,\omega)$ in terms of such components. Following Ref.~\cite{paper052}, we write a solution of the form $\Eb=\Eb^{\rm bulk}+\Eb^{\rm surf}$, where
\begin{align} \label{Eext} 
\Eb_j^{\rm \rm bulk}(\Qb,z,\omega) = \frac{4\pi\ii e}{v\epsilon_j} \ee^{\ii \omega z/v} \frac{\qb-k \epsilon_j \vb/c}{q^2-k^2 \epsilon_j} 
\end{align}
with $q_z =\omega/v$ is the bulk field generated by an electron moving in an infinite medium $j$ (=0 and 1 for vacuum and the material), while
\begin{align} \label{trfield} 
\Eb^{\rm surf}(\Qb,z,\omega) = \begin{cases} 
A\, \ee^{-\ii q_{z0}z} \eb_{\Qb p}^{0-}, &  z<0, \\
\sum_{\pm} C_\pm\, \ee^{\pm\ii q_{z1}z}\, \eb_{\Qb p}^{1\pm}, &  0<z<L, \\
B\, \ee^{\ii q_{z0}z}\, \eb_{\Qb p}^{0+}, &  z>L,
\end{cases}
\end{align}
is the surface-scattered field (with the film contained in the $0<z<L$ region), which consists of outgoing waves in the vacuum regions and a superposition of counter-propagating waves inside the film. For the perpendicular trajectory under consideration, the field is entirely made of p-polarized components with polarization vectors defined as $\eb_{\Qb p}^{j\pm} = \big(\pm q_{zj} \Qb -Q^2 \zz\big)/(k Q\sqrt{\epsilon_j})$ with $q_{zj}=\sqrt{k^2\epsilon_j-Q^2}$. The amplitudes $A$, $B$, and $C_\pm$ are determined from the continuity of the parallel components of the total electric and magnetic fields at $z=0$ and $z=L$ \cite{paper052}, as shown in Appendix \ref{apdxcoefs}.

Inserting Eq.~(\ref{trfield}) into Eqs.~(\ref{eelsfield1}) and (\ref{gammafield}), we find the following expression for the surface contribution to the nonlocal EELS probability:
\begin{widetext}
\begin{align} \label{eelstr} 
\Gamma^{\rm surf}(R,\omega) = \frac{e c}{2\pi^2 v \hbar\omega^2} \int_0^\infty dQ \, Q^2 J_0(Q R)\, \Imm\Bigg\{\frac{A}{\omega/v+q_{z0}}-\frac{B\ee^{-\ii(\omega/v-q_{z0})L}}{\omega/v-q_{z0}}+\sum_{\pm} \frac{C_\pm \big[\ee^{-\ii(\omega/v\mp q_{z1})L}-1\big]}{\sqrt{\epsilon}(\omega/v\mp q_{z1})} \Bigg\},
\end{align}
\end{widetext}
where the induced field has been integrated over $z$. Finally, the decoherence probability is obtained by inserting $\Gamma^{\rm surf}(R,\omega)$ into Eq.~(\ref{decoP}). Incidentally, by using the bulk field given by Eq.~(\ref{Eext}) instead of Eq.~(\ref{trfield}), we recover Eq.~(\ref{bulkan1}) with $L$ identified as the film thickness, since vacuum regions do not contribute to $\Gamma^{\rm bulk}(R,\omega)$. Next, we present numerical results for $P^{\rm surf}$ and $P=P^{\rm bulk}+P^{\rm surf}$.

Examining the frequency-resolved decoherence probability for a $200$~keV electron traversing a $100$~nm-thick film (Fig.~\ref{Fig4}b), we again observe a low-$\omega$ divergence in the limit of infinite path separation. For LiF, this divergence scales as $\omega^{-1}$ at finite temperature and originates from the bulk contribution. In metals, the divergence is stronger, scaling as $\omega^{-1}$ and $\omega^{-2}$ at zero and finite temperature, respectively (see Supplementary Fig.~\ref{FigS3} for results in the zero-temperature limit). Introducing a finite path separation suppresses these divergences, rendering the decoherence probability integrable again. Notably, the logarithmic divergence associated with the bulk contribution in Al and Au (see Fig.~\ref{Fig2}b) is canceled by the surface contribution. While we have fixed $L=100$~nm throughout this section, we explore the dependence on $L$ in Supplementary Fig.~\ref{FigS4}.

Retardation plays an important role in the observed divergences, as illustrated in Fig.~\ref{Fig4}c, where we compare the decoherence probability for an Al film calculated with a finite speed of light to that obtained in the $c \to \infty$ limit (with similar results observed for Au). Retardation enters through transition radiation [i.e., $Q<k$ components of Eq.~(\ref{Eext}) for $j=0$], which dominates the divergence behavior. In contrast, in the nonretarded limit, the divergence originates from the excitation of long-wavelength surface plasmons. Interestingly, although plasmons decay rapidly with distance, partial coherence is retained at finite path separations.

In LiF, we observe low-energy features ($\lesssim0.1$~eV, see Fig.~\ref{Fig4}b) associated with the excitation of phonon polaritons and dominated by guided modes [i.e., $Q>k$ components in Eq.~(\ref{trfield})], as we show in Fig.~\ref{Fig4}d by comparing results for $P^{\rm surf}$ performing the full integral over $Q$ with those obtained by only integrating above $k$. In the $\omega\to0$ limit, $P^{\rm surf}$ remains finite even at infinite path separation, showing that the divergence in $P$ for LiF (Fig.~\ref{Fig4}b) originates exclusively from the bulk contribution.

For metals, the $\omega^{-2}$ divergence becomes increasingly more pronounced at larger $d_\perp$, where wavelengths comparable to the path separation enhance decoherence. The comparison of surface and bulk contributions in Fig.~\ref{Fig4}e shows that the latter remains dominant even for $d_\perp\approx 50\,{\rm \mu m}$, but the relative weight of $P^{\rm surf}$ increases with $d_\perp$, a trend that becomes more apparent at elevated temperatures. Consequently, metals display an enhanced temperature dependence, with $P$ varying by $>1\%$ when moving from $T=0$ to $T=1000$~K (Fig.~\ref{Fig4}f). In contrast, for LiF, the surface contribution associated with phononic modes is reinforced at elevated temperatures, but does not produce a divergence as pronounced as in metals, while the high-energy bulk contribution still dominates (Fig.~\ref{Fig4}e), leading to a similar fractional change of $P$ with temperature as in metals.

\begin{table*}
\includegraphics[width=0.75\textwidth]{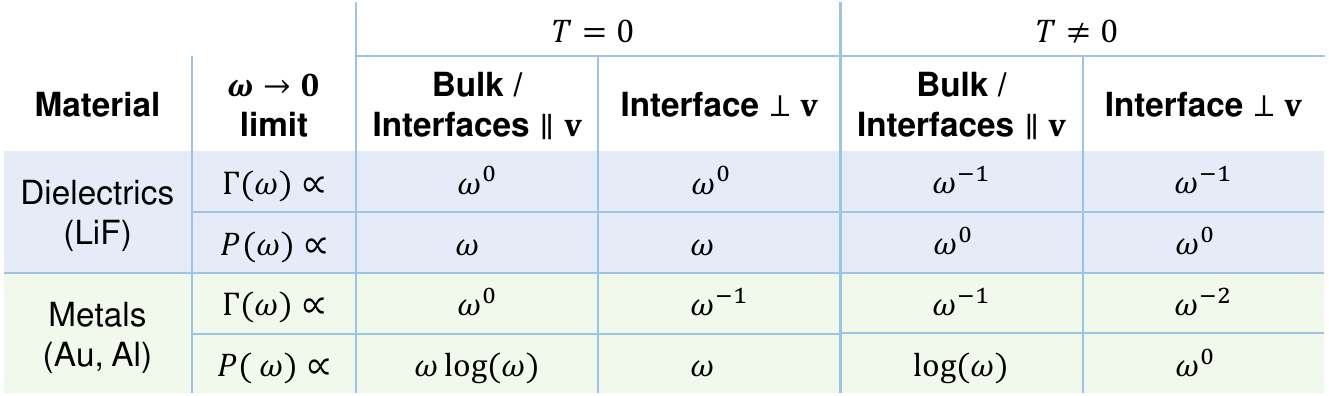}
\caption{{\textbf{Low-frequency behavior across different materials and geometries.} In all configurations discussed within this work, the EELS probability $\Gamma$ diverges at least as $\omega^{-1}$ in the low-frequency limit at finite temperatures. This behavior is inherited from the Bose--Einstein distribution $n_T(\omega\to0)\propto\omega^{-1}$. The low-frequency divergences are regularized in the frequency-resolved decoherence probability $P(\omega)$, yielding finite $\omega$-integrated values at finite interpath separations $d_\perp$}.}
\label{Table1}
\end{table*}

\begin{figure*}
\centering\includegraphics[width=1.0\textwidth]{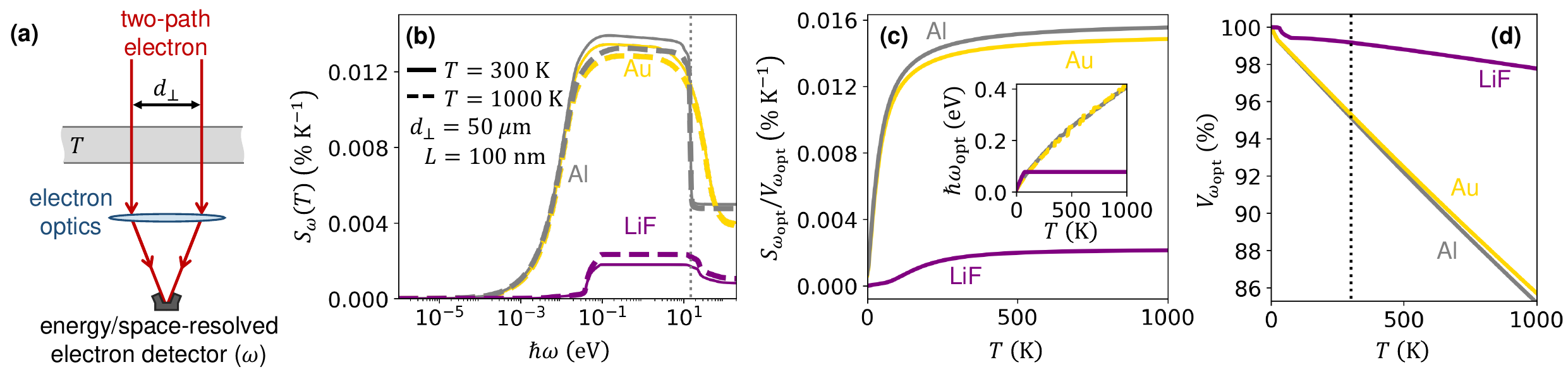}
\caption{\textbf{Temperature sensing through decoherence}.
\textbf{(a)}~A normally incident e-beam prepared in a two-path superposition intersects a thin film placed at a temperature $T$. The transmitted e-beam is energy- and space-resolved at a detector, where the two paths are recombined. The spectrally resolved interference is used to measure the temperature of the sample.
\textbf{(b)}~Energy-filtered temperature sensitivity normalized to the corresponding fringe visibility as a function of energy-loss cutoff for $200$~keV electrons traversing Al, Au, or LiF films of thickness $L=100$~nm at $T=300$~K or $T=1000$~K. We set the path separation to $d_\perp = 50\,{\rm \mu m}$. The dotted line indicates the bulk plasma frequency of Al.
\textbf{(c)}~Temperature dependence of the energy-filtered temperature sensitivity for a $T$-dependent optimum energy-loss cutoff $\hbar\omega_{\rm opt}(T)$ (see inset).
\textbf{(d)}~Fringe visibility under the conditions of panel (c). The dotted line indicates $T=300$~K.}
\label{Fig5}
\end{figure*}

\section{Temperature dependence}
\label{sectemp}

The strong temperature dependence of electron decoherence suggests its use for nanoscale thermometry. As shown above, the dependence is dramatic for low-frequency excitations ($\hbar\omega\lesssim\kB T$). However, when integrating over all frequencies, contributions from higher-energy modes often dominate, reducing the relative weight of the temperature dependence. The temperature sensitivity can be increased by suppressing these high-energy contributions through energy filtering of the e-beam after interaction, retaining only low-frequency excitations. Under these conditions, it is useful to introduce an energy-filtered decoherence probability 
\begin{align} \nonumber 
\tilde{P}_\omega(\Rb,\Rb') = \int_0^{\omega}d\omega'\; P(\Rb,\Rb',\omega'),
\end{align}
which quantifies the decoherence associated with electrons that have gained or lost energy up to $\hbar\omega$ during the interaction. This quantity is particularly relevant in experiments employing energy-filtered e-beams \cite{H05,EDP07}, as it allows us to identify those inelastic channels that contribute maximally to electron decoherence. In practice, $\tilde{P}_\omega(\Rb,\Rb')$ can be readily measured by inserting an energy filter between the interaction region and the electron detector or, ideally, by using a spectrometer. The filter could be introduced at any point after the interaction region, for example, just before the object plane $z = z_o$ (see Fig.~\ref{Fig1}b). However, it may be more convenient to place it after beam recombination, for instance, after the second lens of the microscope. If a spectrometer is used, interferograms can be recorded along one axis of the camera, while the other axis resolves the electron kinetic energy. In this configuration, energy-filtered visibilities can be obtained by integrating the interferograms over a selected final energy range. For illustration, we focus on a two-path electron traversing a thin film (Fig.~\ref{Fig5}a), as this configuration enhances temperature sensitivity owing to the higher-order divergence in the loss probability (see Sec.~\ref{secperp} and Table~\ref{Table1}).

To quantify the thermometric performance, we introduce the temperature sensitivity function, defined as the derivative of the fringe visibility with respect to temperature for the energy-filtered beam,
\begin{align} \nonumber 
S_\omega(T)=\bigg|\frac{\partial V_\omega}{\partial T} \bigg|=\ee^{-\tilde{P}_\omega} \bigg|\frac{\partial \tilde{P}_\omega}{\partial T} \bigg|,
\end{align}
where we use the expression $V_\omega=\ee^{-\tilde{P}_\omega(d_\perp)}$ for the visibility [see Eq.~(\ref{intdemo})]. To improve performance, we choose a cutoff frequency $\omega$ such that strong high-frequency features (e.g., plasmons in metals) are suppressed, while maintaining $\tilde{P}_\omega$ sufficiently large to remain measurable. Figure~\ref{Fig5}b shows $S_\omega$ for Al, Au, and LiF films of $100$~nm thickness. For all materials, within an energy-loss window between $100$~meV and $10$~eV, $S_\omega$ reaches significantly higher values than those achieved when integrating over the full energy range. The lower bound to this energy window is determined by the interpath distance $d_\perp=50\,\mu$m, which imposes an effective cutoff on the wavelength of modes that can effectively imprint which-path information onto the environment and, thus, deplete interpath coherence. Interestingly, a large temperature sensitivity is observed for LiF, which is associated with phonon losses at $\sim 0.06$~eV (i.e., close to $\kB T$). In conclusion, the sensitivity is optimized for energy cutoffs above optical phonons, but below the electronic band gap.

To formalize this idea, we define an optimal cutoff frequency $\omega_{\rm opt}(T)$ as the value of $\omega$ that maximizes the temperature sensitivity $S_{\omega}(T)$ at a fixed temperature $T$. As shown in Fig.~\ref{Fig5}c, the maximum sensitivity varies mildly for $T\gtrsim200$~K and coincides with the maxima identified in Fig.~\ref{Fig5}b, predicting a nearly linear dependence of visibility on temperature. Indeed, as shown in Fig.~\ref{Fig5}d, the visibility exhibits an approximately linear dependence on $T$ over most of the range considered, consistent with the low-frequency approximation $n_T(\omega)\approx \kB T/\hbar\omega$ together with the small-decoherence limit $V_{\omega} \approx 1-\tilde{P}_\omega$. At $\omega=\omega_{\rm opt}$, the visibility for metals and LiF changes by 10\% and 2\%, respectively. Finally, we note that increased sensitivity to low temperatures can be gained by increasing the interpath separation $d_\perp$ in combination with a reduced energy cutoff.

\begin{table*}[t]
\includegraphics[width=0.75\textwidth]{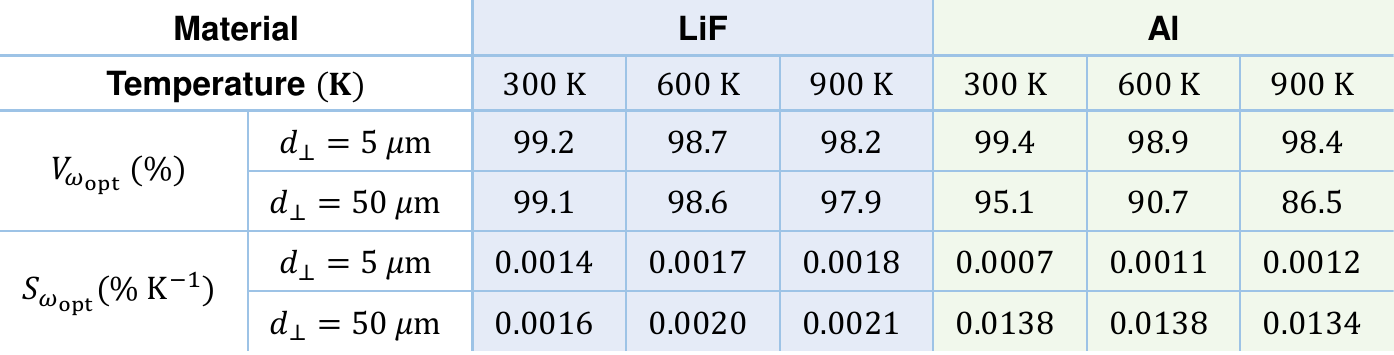}
\caption{\textbf{Fringe visibility and sensitivity to decoherence under different configurations.} We provide representative values of the visibility $V$ and sensitivity $S$ for different path separations $d_\perp$ and temperatures under the conditions of Fig.~\ref{Fig5} for LiF and Al films of 100~nm thickness. Results are presented for a $T$-dependent energy-loss cutoff $\hbar\omega_{\rm opt}$ as described in Fig.~\ref{Fig5}.}
\label{Table2}
\end{table*}

\section{Conclusions and outlook}
\label{conclusions}

In summary, we present a comprehensive study of the loss of spatial coherence experienced by free electrons due to interactions with material structures, which produce an impact-parameter-dependent entanglement of an e-beam with different excitations (i.e., a dependence on lateral position). Results are offered for two-path e-beams propagating inside a homogeneous material or near planar surfaces under parallel and perpendicular orientations. Decoherence should manifest in a loss of visibility upon recombination of the two paths, as illustrated above for a specific implementation suitable for electron microscopes. Different types of materials (e.g., plasmonic and phonon-polaritonic), spectral regions, and geometries are analyzed based on a unified formalism, offering a broad perspective of material-driven electron decoherence for general application in electron microscopy studies.

As a general rule, by recording all scattered electrons, we find that bulk excitations, such as bulk plasmons and interband transitions, play a dominant role in the depletion of coherence for metallic and dielectric materials, respectively, with only a weak dependence on temperature in the $0-1000$~K range. This is in contrast to the contribution of surface modes. The temperature dependence of decoherence is governed by coupling to thermally populated low-frequency modes, such as free radiation, optical guided modes (i.e., Cherenkov emission), and long-lived surface plasmon polaritons: the population of those modes is enhanced by the infrared divergence in the Bose--Einstein distribution, which translates into a divergent EELS probability. This divergence is regularized in the decoherence probability, which incorporates the indistinguishability of the two paths when their separation is small compared to the wavelengths associated with the mode fields.

By examining different interaction configurations, we illustrate the roles played by various types of modes in the loss of coherence. For a split beam propagating entirely inside a bulk material at finite temperature, the emission of Cherenkov radiation leads to a non-divergent coupling at low frequencies, which is, however, rendered divergent by the thermal population at finite temperature. Interestingly, when the two electron paths propagate parallel to a planar material surface, decoherence is enhanced if one of the beams moves inside the medium and the other in the surrounding vacuum. This effect becomes particularly pronounced for materials such that the internal path couples to localized bulk excitations and the external path to surface modes. In this configuration, we find order-unity decoherence probabilities for relatively short electron path lengths (e.g., $50$~nm in Al), implying a dramatic reduction in fringe visibility when the two paths are interfered. A low-frequency behavior of the loss probability analogous to that of the bulk configuration arises from both Cherenkov radiation inside the material and its radiative analogue outside.

We observe a more complex interplay of decoherence mechanisms when the two beams traverse a material film perpendicularly. In this configuration, the broad range of in-plane and out-of-plane momenta accessible to excitations multiplexes the inelastic channels responsible for decoherence by launching surface and waveguided modes and generating transition radiation. For conductive samples, this geometry produces an additional $\omega^{-1}$ divergence in the energy-loss probability at zero temperature. At nonvanishing temperatures, this becomes a $\omega^{-2}$ divergence, again regularized in the decoherence probability (see Table~\ref{Table1}). Overall, this configuration exhibits the strongest temperature dependence among the studied geometries. For example, while the variation of the decoherence probability $P$ over the $0-1000$~K range for Al films of 100~nm thickness is on the order of $1\%$ of its $T=0$ value, the relative variation inside a bulk material is nearly one order of magnitude smaller (see Table~\ref{Table2}). In LiF films, there is no zero-temperature divergence, and guided-mode excitations near the bulk phonon resonance in the meV range give rise to a modest temperature dependence.

Another important consequence of a higher-order divergence is the enhanced interpath dependence of decoherence, stemming from its connection to the cutoff wavelength at which the two paths become indistinguishable. In this context, we find that, for Al, the loss of interference fringes increases linearly with the separation between the two paths, whereas LiF films display a slower, logarithmic increase. This behavior is clearly captured by the ratio between surface and bulk contributions to decoherence (see Fig.~\ref{Fig4}e), as well as by the corresponding values of visibility and temperature sensitivity reported in Table~\ref{Table2}. These findings are consistent with those predicted for a semi-infinite perfect-conductor plate \cite{paper425}, where an inverse-frequency spectral divergence yields a logarithmic increase in decoherence with path separation, while an inverse-square dependence results in linear growth.

The strong temperature dependence of the low-frequency energy-loss region in the perpendicular e-beam configuration suggests the study of decoherence as an approach to temperature sensing. To maximize this dependence, energy-filtered interference patterns can be recorded, allowing us to suppress temperature-independent contributions arising from bulk plasmons and electron-hole excitations in the $1-50$~eV spectral region. We show that the spectral filtering window can be tuned to optimize temperature sensitivity. We find that collecting electrons with energies at distances from the zero-loss peak well below the plasmon energy yields an interference visibility of $\sim97$\%, which in turn results in measurable changes of order $0.1$\% near room temperature across a $6$~K variation in an Al film.

State-of-the-art nanoscale thermometry in transmission electron microscopy achieves combined thermal and spatial resolutions on the order of K-nm \cite{MHW15,CLA20,LB18,ILF18,MSL21,POK24,PLD25}, with recent extensions to nanosecond temporal resolution at the cost of reduced thermal precision \cite{CAB25}. However, these approaches are typically material-specific and rely on specialized instrumentation. For instance, methods based on plasmon energy shifts \cite{MHW15,CLA20,PLD25,CAB25} track lattice thermal expansion, limiting their applicability to materials with well-defined plasmonic features and sizable thermal expansion coefficients. Techniques based on phonon population ratios in energy-loss and energy-gain spectra \cite{LB18,ILF18,CAB25} require highly monochromated e-beams with energy spreads of only a few tens of meV. Cathodoluminescence-based methods further depend on temperature-sensitive emission features and an additional optical collection module \cite{MSL21,POK24}.

In contrast, our proposed approach applies to any conductive sample, as it exploits the high coupling probability to low-frequency radiative and surface plasmon modes that arise for penetrating electron trajectories at dielectric--conductor interfaces. The required instrumentation is limited to energy filtering as well as a sufficient transverse spatial coherence to enable coherent beam splitting into two paths separated by tens of microns. We also require structure sizes larger than the path-to-path separation to support the required long-wavelength modes. Under these conditions, temperature sensitivities on the order of $0.1\%$~K$^{-1}$ (see Table~\ref{Table2}) can be achieved with a relatively weak material dependence (see Fig.~\ref{Fig5}).

Beyond potential applications, the theoretical formulation of decoherence probabilities in terms of the electromagnetic Green tensor and the spectral decomposition of energy-loss channels provides a natural framework for generalization to more complex geometries, combining analytical insight with numerical methods. While our analysis focuses on electrons prepared in a spatial superposition of two point-like paths in the transverse plane, the formalism can be straightforwardly extended to configurations in which initially collimated beams are modified by a combination of elastic and inelastic scattering processes, leading to distortions of the full spatial structure of the electron density matrix (see Appendix~\ref{freepropagation}). In such scenarios, decoherence manifests not only as a reduction of fringe visibility but, more generally, as a redistribution of coherence across spatial modes. This perspective opens the door to the use of advanced reconstruction algorithms aimed at retrieving the reduced electron density matrix from experimentally accessible observables such as diffraction intensities \cite{CKL16}. In particular, ptychographic approaches \cite{LLM25} adapted to partially coherent e-beams \cite{COJ20,XNS24} offer a promising route to reconstruct both the amplitude and phase of the electron density matrix, including its off-diagonal elements, which directly encode decoherence. Such reconstruction schemes could enable quantitative extraction of temperature-dependent decoherence signatures, thereby providing an alternative and potentially powerful route to nanoscale thermometry based on full density-matrix reconstruction. 

\section*{ACKNOWLEDGEMENTS} 
We thank Archie Howie and Jo Verbeeck for stimulating and enjoyable discussions.
This work has been supported in part by the European Research Council (Adv. Grant 101141220-QUEFES), the European Commission (FET-Proactive 101017720-eBEAM), the Spanish MICIU (PID2024-157421NB-I00 and Severo Ochoa CEX2024-001490-S), and the CERCA Program.

\appendix

\section{Derivation of the intensity profile in Fig.~\ref{Fig1}} \label{apdxprop}

To describe the propagation of an e-beam in the configuration shown in Fig.~\ref{Fig1}b, we assume the two paths to be wide enough as to neglect diffraction over the entire setup (see main text). In addition, we assume axially-symmetric Gaussian profiles, such that each path $j=A,B$ has an associated transverse wavefunction $\psi_\perp(\Rb-\Rb_j)$, where
\begin{align} \label{psij0} 
\psi_\perp(\Rb)=\frac{\sqrt{2}}{\pi^{3/4}w_0}\ee^{-R^2/w_0^2}
\end{align}
is an intensity-normalized component with a small Gaussian radius $w_0$ compared to the interpath separation $|\Rb_A-\Rb_B|$. The initial electron density matrix before interaction is $\rho(\Rb,z,\Rb',z)=(1/2)\sum_{j,j'=A,B}\psi_\perp(\Rb-\Rb_j)\psi_\perp(\Rb'-\Rb_{j'})$. Note that the complete sub-beam wavefunctions depend on the longitudinal coordinate $z$ through a propagation phase, which disappears in the product inside the density matrix when this is evaluated at the same $z$ plane.

The interaction is introduced through Eq.~(\ref{evol}). We assume that neither the decoherence probability $P$ nor the elastic phase $\chi$ varies significantly across the transverse extent of the two paths, and thus, they can be evaluated at the path centers as
\begin{align} \nonumber 
&\ee^{-P(\Rb,\Rb')+\ii\chi(\Rb,\Rb')} \psi_\perp(\Rb-\Rb_j,z_0)\psi_\perp(\Rb'-\Rb_{j'},z_0) \approx \\ 
&\ee^{-P(\Rb_j,\Rb_{j'})+\ii\chi(\Rb_j,\Rb_{j'})} \psi_\perp(\Rb-\Rb_j,z_0)\psi_\perp(\Rb'-\Rb_{j'},z_0). \nonumber
\end{align}
Under this approximation, the entire beam propagation can be reduced to the independent evolution of each path. After the interaction, the corresponding density matrix becomes
\begin{align} \label{effprop} 
\rho(\Rb,z,\Rb',z) = &\frac{1}{2}\sum_{j,j'=A,B}\psi_\perp(\Rb-\Rb_j)\psi_\perp(\Rb'-\Rb_{j'}) \nonumber\\
&\times \ee^{-P(\Rb_j,\Rb_{j'})+\ii\chi(\Rb_j,\Rb_{j'})},
\end{align}
which includes path-dependent interaction factors.

We now introduce an electrostatic deflection of the sub-beams, which brings them together at an overlapping region in a plane $z=z_o$ placed a distance $d_\parallel$ from the mirror (see Fig.~\ref{Fig1}b). Assuming a small deflection angle (i.e., $d_\parallel\gg R_j$) that maintains paraxial propagation conditions, the sub-beam wavefunctions become $\ee^{-\ii q_0\Rb_j\cdot\Rb/d_\parallel}\psi_\perp(\Rb)$ at the $z_o$ plane, up to a global position-independent phase factor. In addition, a two-lens microscope is introduced to produce an enlarged image of the $z_o$ plane at the conjugate, electron-detector plane $z_i$. For the parameters shown in Fig.~\ref{Fig1}b, the magnification factor is given by the focal-length ratio $g=f_2/f_1$, under the assumption that the lenses are separated by a distance $f_1+f_2$ and the condition $g^2=-(d_1-f_1)/(d_2-f_2)$ is fulfilled, where $d_1$ and $d_2$ are the distances of the $z_o$ and $z_i$ planes to lenses 1 and 2, respectively. The sub-beam wavefunctions at $z_i$ become $\propto\ee^{\ii\Qb_j\cdot\Rb}\psi_\perp(-\Rb/g)$, where $\Qb_j=(q_0/gd_\parallel)\Rb_j$ \cite{paper451}. Performing a separate propagation of each path through this system and considering the trace of the density matrix [Eq.~(\ref{effprop})], the detected electron intensity is given by $I(\Rb)\propto\rho(\Rb,z_i,\Rb,z_i)=(1/2)\,\big|\psi_\perp(-\Rb/g)\big|^2$ $\times\sum_{j,j'}\ee^{\ii(\Qb_j-\Qb_{j'})\cdot\Rb}\ee^{-P(\Rb_j,\Rb_{j'})+\ii\chi(\Rb_j,\Rb_{j'})}$. Finally, insering Eq.~(\ref{psij0}) into this expression, we readily obtain Eq.~(\ref{intdemo}).

\section{Lindhard and Mermin dielectric functions} \label{apdxeps}

We describe nonlocal effects in bulk metals by using the Lindhard dielectric function obtained in the RPA for a free-electron gas \cite{L1954,DG02}. For longitudinal fields, it is given by \cite{DG02}
\begin{align} \nonumber 
&\epsilon^{\rm L}_{\rm lon}(q,\omega) \\
&= 1+ \frac{2\me e^2 k_F}{\pi \hbar^2 q^2} \bigg[1&+R\bigg(\frac{q}{2k_F},\frac{\omega}{\varepsilon_F}\bigg)
+R\bigg(\frac{q}{2k_F},-\frac{\omega}{\varepsilon_F}\bigg)\bigg] \nonumber
\end{align}
where $\hbar\varepsilon_F = \hbar^2 k_F^2/2\me$ is the Fermi energy, $k_F$ is the Fermi wavenumber, and
\begin{align} \nonumber 
R(x,y) = \frac{1}{2x}\bigg[1-\frac{(x^2+y)^2}{4x^2}\bigg]\log\bigg(\frac{x^2+2x+y}{x^2-2x+y}\bigg).
\end{align}
An inelastic damping rate $\eta$ cannot be simply introduced in this dielectric function through the $\omega\to\omega+\ii\eta$ substitution because it breaks the conservation of local electron density. Instead, we adopt Mermin's prescription to enforce such conservation, leading to the dielectric function \cite{M1970}
\begin{align} \nonumber 
\epsilon^{\rm M}_{\rm lon}(q,\omega) = 1+\frac{(\omega+\ii \eta) [\epsilon_{\rm L}(q,\omega+\ii\eta)-1]}{\omega+\ii\eta [\epsilon_{\rm L}(q,\omega+\ii\eta)-1]/[\epsilon_{\rm L}(q,0)-1]}.
\end{align}
In the local approximation, the Mermin dielectric function reduces to the Drude model [$\epsilon^{\rm M}_{\rm lon}(q\to 0,\omega)=\epsilon^{\rm D}(\omega)$, see Eq.~(\ref{drude})]. For the transverse component, we also use the Drude model, as nonlocal corrections are negligible for metals such as Al and Au \cite{DG02}.

\section{Induced field for an electron crossing a thin film} \label{apdxcoefs}

As mentioned in Sec.~\ref{secperp}, the coefficients for the induced field in Eq.~(\ref{trfield}) are obtained by imposing the continuity of the parallel components of the electric and magnetic fields at the $z=0$ and $z=L$ surface planes. For the electric field, we directly use Eqs.~(\ref{Eext}) and~(\ref{trfield}). For the magnetic field, we obtain the {\it bulk} form in a homogeneous medium $j$ (i.e., either the material or the external vacuum) by applying Faraday's law to Eq.~(\ref{Eext}), leading to
\begin{align} \nonumber 
{\bf H}^{\rm bulk}(\Qb,z,\omega) = \frac{4\pi\ii e Q \ee^{\ii \omega z/v}}{c(q^2-k^2\epsilon_j)}\,\, \eb_{\Qb s}
\end{align}
with $\eb_{\Qb s}=(-q_y\xx+q_x\yy)/Q$, while the surface contribution is obtained from Eq.~(\ref{trfield}) as
\begin{align} \nonumber 
&{\bf H}^{\rm surf}(\Qb,z,\omega) \\
&= \eb_{\Qb s}\;\begin{cases} 
A\, \ee^{-\ii q_{z0}z}, &\quad  z<0, \\
\sqrt{\epsilon_j}\sum_{\pm}\,C_\pm\, \ee^{\pm\ii q_{z1}z}, &\quad  0<z<L, \\
B\, \ee^{\ii q_{z0}z}, &\quad  z>L.
\end{cases} \nonumber
\end{align}
Note that both bulk and surface magnetic fields are $s$-polarized. Projecting the electric and magnetic fields over $\Qb$ and $\eb_{\Qb s}$, respectively, yields a linear system of equations for the unknown coefficients $A$, $B$, and $C_\pm$, which can be written in compact matrix form as $M\cdot {\bf C} = {\bf b}$, where ${\bf C} = (A,B,C_+,C_-)^{\intercal}$, 
\begin{align} \nonumber 
M = 
\begin{pmatrix}
1 & 0 & -\sqrt{\epsilon} & -\sqrt{\epsilon} \\
0 & -\ee^{\ii q_{z0}L} & \xi_1\sqrt{\epsilon} & \xi_1^{-1}\sqrt{\epsilon} \\
q_{z0}/k & 0 & q_{z1}/k\sqrt{\epsilon} & -q_{z1}/k\sqrt{\epsilon} \\
0 & \xi_0q_{z0}/k & -\xi_1q_{z1}/k\sqrt{\epsilon} &  \xi_1^{-1}q_{z1}/k\sqrt{\epsilon}
\end{pmatrix},
\end{align}
and ${\bf b} = (b_1,-b_1 \ee^{\ii \omega L/v},b_2,-b_2 \ee^{\ii\omega L/v})^\intercal$, with
$\xi_j=\ee^{\ii q_{zj}L}$,
$b_1=C_0\,(\epsilon-1)$,
$b_2=C_0\,(1/\epsilon-1)\,(c/v)\,\big(1+\epsilon-q^2/k^2\big)$,
and $C_0=4\pi \ii e Q(k^2/c)/\big[(q^2-k^2)(q^2-k^2\epsilon)\big]$.
The solution ${\bf C} = M^{-1} \cdot{\bf b}$ is explicitly given by
\begin{align} \nonumber 
&C_+=C_0\,(\epsilon-1)\,t_p^{01}\,\big[\kappa^2_{+} + \ee^{\ii\omega L/v}\xi_1\, r_p^{01} \kappa_-^2\big] \,F^{-1}, \\
&C_-=\ee^{2\ii q_{z1} L}\,C_0\,(1-\epsilon)\,t_p^{01}\,\big[  \kappa^2_{-} \ee^{\ii\omega L/v}\xi_1^{-1} + r_p^{01}  \kappa_+^2\big] \,F^{-1}, \nonumber\\
&A=\sqrt{\epsilon}\, (C_++C_-) +C_0(\epsilon-1), \nonumber\\
&B=\xi_0^{-1}\big\{\sqrt{\epsilon}\, \big[C_+ \,\xi_1 +C_- \,\xi_1^{-1} \big]
+C_0(\epsilon-1) \,\ee^{\ii\omega L/v}\big\}, \nonumber
\end{align}
where $F=2\epsilon (vk/c)\,q_{z0} \,\big[1-(r_p^{01}\xi_1)^2\big]$ and $\kappa_{\pm}^2 = q^2-k^2[1+\epsilon\pm(q_{z0}v/\omega)\epsilon]$.

\section{Propagation of arbitrary transverse e-beam profiles} \label{freepropagation}

Although we focus the present work on two-path e-beams for tutorial purposes, the formalism in Sec.~\ref{sectheory} can be applied to any transverse profile. Propagation from a post-interaction plane $z=z_0$ to a subsequent plane $z=z_1$ along the e-beam column simply mixes different transverse components of the density matrix in Eq.~(\ref{evol}). By moving from transverse coordinates $\Rb$ and $\Rb'$ to transverse momentum space $\Qb$ and $\Qb'$ according to $\rho(\Qb,z_0,\Qb',z_0) = \int d^2\Rb\int d^2\Rb' \ee^{-\ii (\Qb\cdot \Rb-\Qb'\cdot\Rb')}\rho(\Rb,z_0,\Rb',z_0)$, the propagated state can be written
\begin{align} \nonumber 
\rho(\Rb,z_1,&\Rb',z_1)=\int  \frac{d^2\Qb}{(2\pi)^2} \int\frac{d^2\Qb'}{(2\pi)^2} \, \rho(\Qb,z_0,\Qb',z_0) \\
&\times  \ee^{\ii (\Qb\cdot \Rb-\Qb'\cdot\Rb')+\ii (q_z-q_z')(z_1-z_0)}, \label{propdens}
\end{align}
where $q_z = \sqrt{q_0-Q^2}$ and $q_z'=\sqrt{q_0-Q^{\prime \,2}}$ are longitudinal electron wavevector components and $q_0$ is the total wavenumber. Under the paraxial approximation ($q_z\approx q_0-Q^2/2q_0$), the $\Qb$ and $\Qb'$ integrals can be performed analytically to recast Eq.~(\ref{propdens}) into a convolution as \cite{paper451}
\begin{align} \nonumber 
\rho(\Rb,z_1,&\Rb',z_1)=\frac{q_0^2}{4\pi^2(z_1-z_0)^2} \\
&\times\int d^2\Rb'' \int d^2\Rb''' \, \rho(\Rb'',z_0,\Rb''',z_0) \nonumber\\
&\quad\quad\quad \times  \ee^{\ii q_0(|\Rb-\Rb''|-|\Rb'-\Rb'''|)/2(z_1-z_0)}. \nonumber
\end{align}
Combining this result with Eq.~(\ref{evol}) explicitly demonstrates how the decoherence probability and the elastic phase jointly influence the evolution of the electronic state. In particular, for a fully coherent incident e-beam characterized by a transverse wavefunction $\psi_\perp(\Rb,z_0)$, the post-interaction densitry matrix is given by $\rho(\Rb,z_0,\Rb',z_0)=\ee^{- P(\Rb,z_0,\Rb',z_0)}\, \ee^{\ii \chi(\Rb,z_0,\Rb',z_0)}\,\psi_\perp(\Rb,z_0)\psi_\perp^*(\Rb',z_0)$.


%

\clearpage
\pagebreak \onecolumngrid \section*{SUPPLEMENTARY FIGURES}
\renewcommand{\thefigure}{S\arabic{figure}} 

\begin{figure*}[h!]
\includegraphics[width=0.66\linewidth]{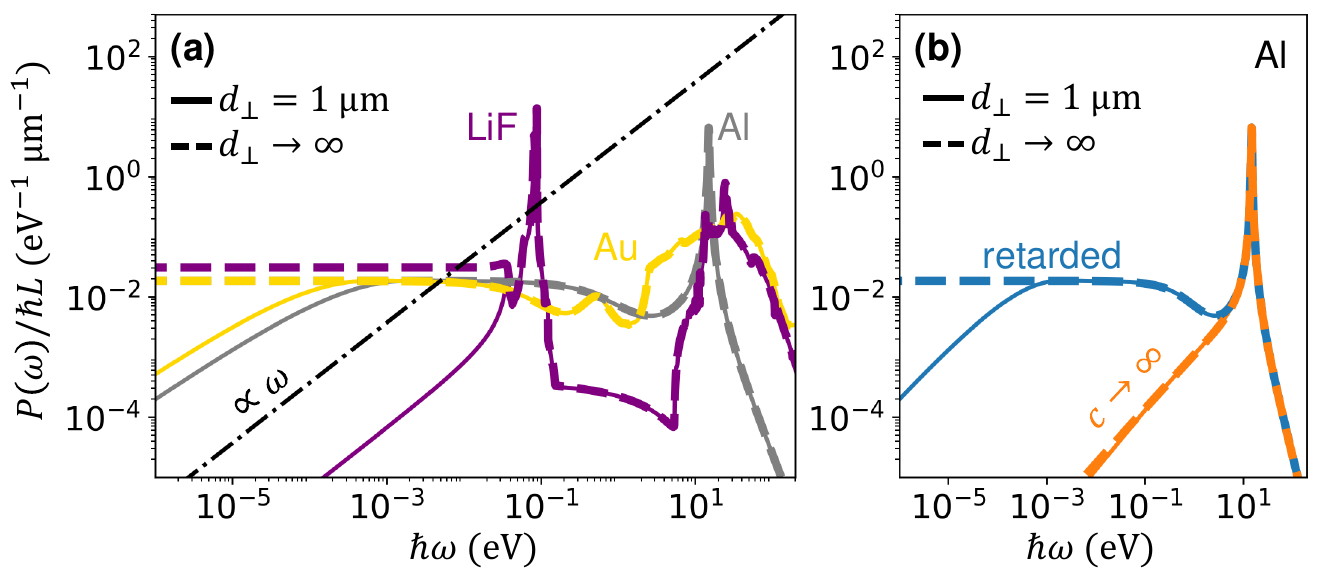}
\caption{\textbf{Zero-temperature limit: Bulk contribution to electron decoherence} \textbf{(a)}~Frequency-resolved decoherence probability $P(\omega)$ for a $200$~keV electron under the configuration of Fig.~2a in the main text when the material is Au, Al, or LiF at $T=0$. We consider $d_\perp=1$~$\mu$m (solid curves) and $d_\perp\to\infty$ (dotted curves, coinciding with the EELS probability for a single-path electron).
\textbf{(b)}~Comparison between the decoherence probability obtained for Al under the configuration in Fig.~2a of the main text at $T=0$ with inclusion of retardation [taken from (a)] and the nonretarded calculation (i.e., taking $c\to\infty$).}
\label{FigS1}
\end{figure*}

\begin{figure*}[h!]
\includegraphics[width=0.33\linewidth]{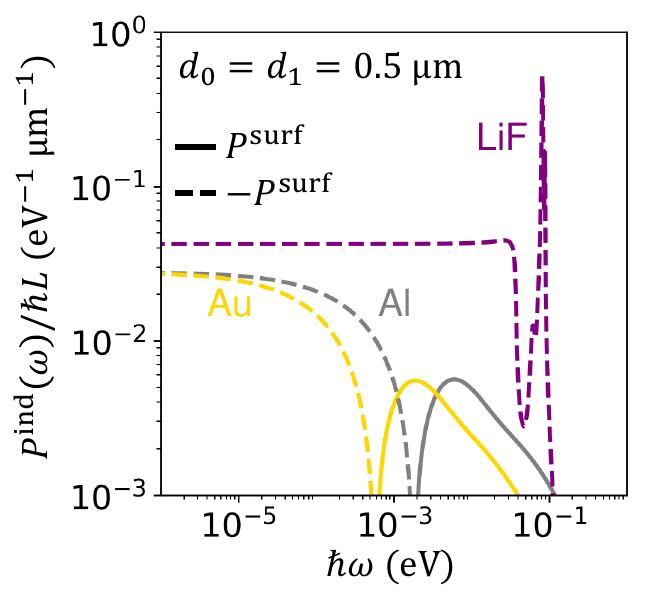}
\caption{\textbf{Zero-temperature limit: Surface contribution to electron decoherence: Parallel e-beam configuration} Surface contribution to the frequency-resolved decoherence probability $P^{{\rm surf}}(\omega)$ for a $200$~keV electron under the configuration of Fig.~3a in the main text when the material is Au, Al, or LiF at $T=0$. We set $d_0=d_1=0.5~\mu$m. As $P^{{\rm surf}}$ becomes negative for some frequencies, we plot both $P^{\rm surf}$ (solid curves) and $-P^{{\rm surf}}$ (broken curves).}
\label{FigS2}
\end{figure*}

\begin{figure*}[h!]
\centering
\includegraphics[width=0.9\linewidth]{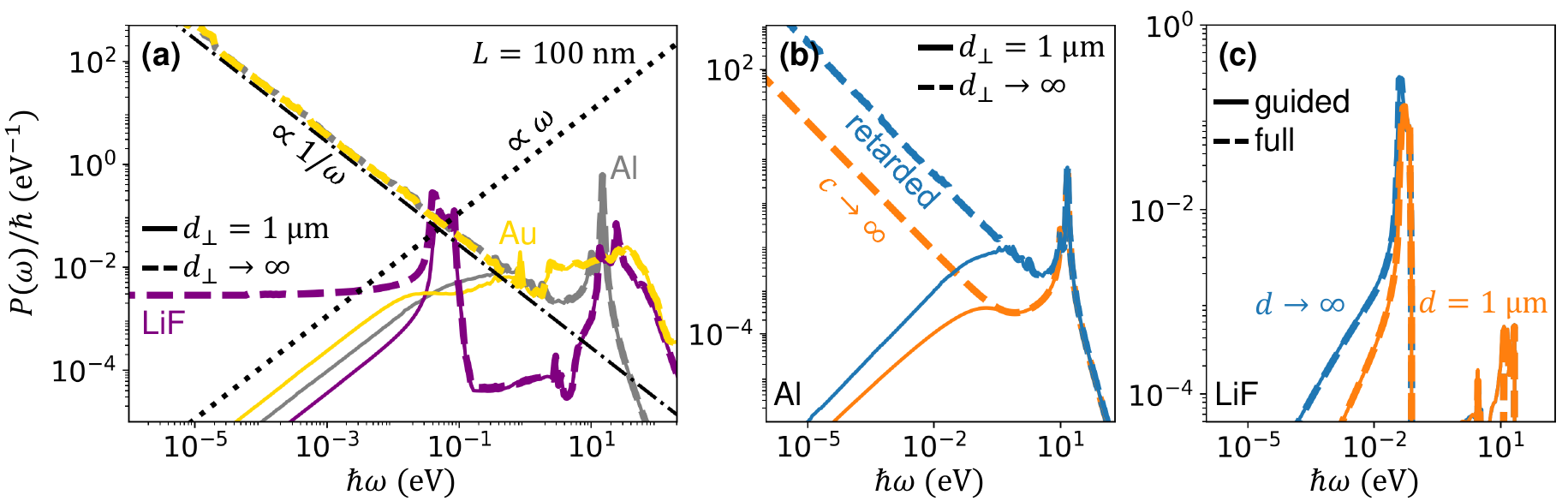}
\caption{\textbf{Zero temperature limit: Surface contribution to electron decoherence: Perpendicular e-beam configuration.} \textbf{(a)}~Frequency-resolved decoherence probability $P(\omega)$ at $T=0$ for a $200$~keV electron under the configuration of Fig.~4a in the main text when the material is Au, Al, or LiF. We consider $d_\perp=1$~$\mu$m (solid curves) and $d_\perp\to\infty$ (dotted curves, coinciding with the EELS probability for a single-path electron).
\textbf{(b)}~Comparison between the decoherence probability at $T=0$ obtained for Al under the configuration in Fig.~4a of the main text with inclusion of retardation [taken from (a)] and the nonretarded calculation (i.e., taking $c\to\infty$). 
\textbf{(c)}~Full surface contribution to the frequency-resolved decoherence probability $P^{\rm surf}(\omega)$ (solid curves) compared with the partial contribution to this quantity arising from guided modes (dashed curves) at $T=0$.}
\label{FigS3}
\end{figure*}

\begin{figure*}[h!]
\centering
\includegraphics[width=0.95\linewidth]{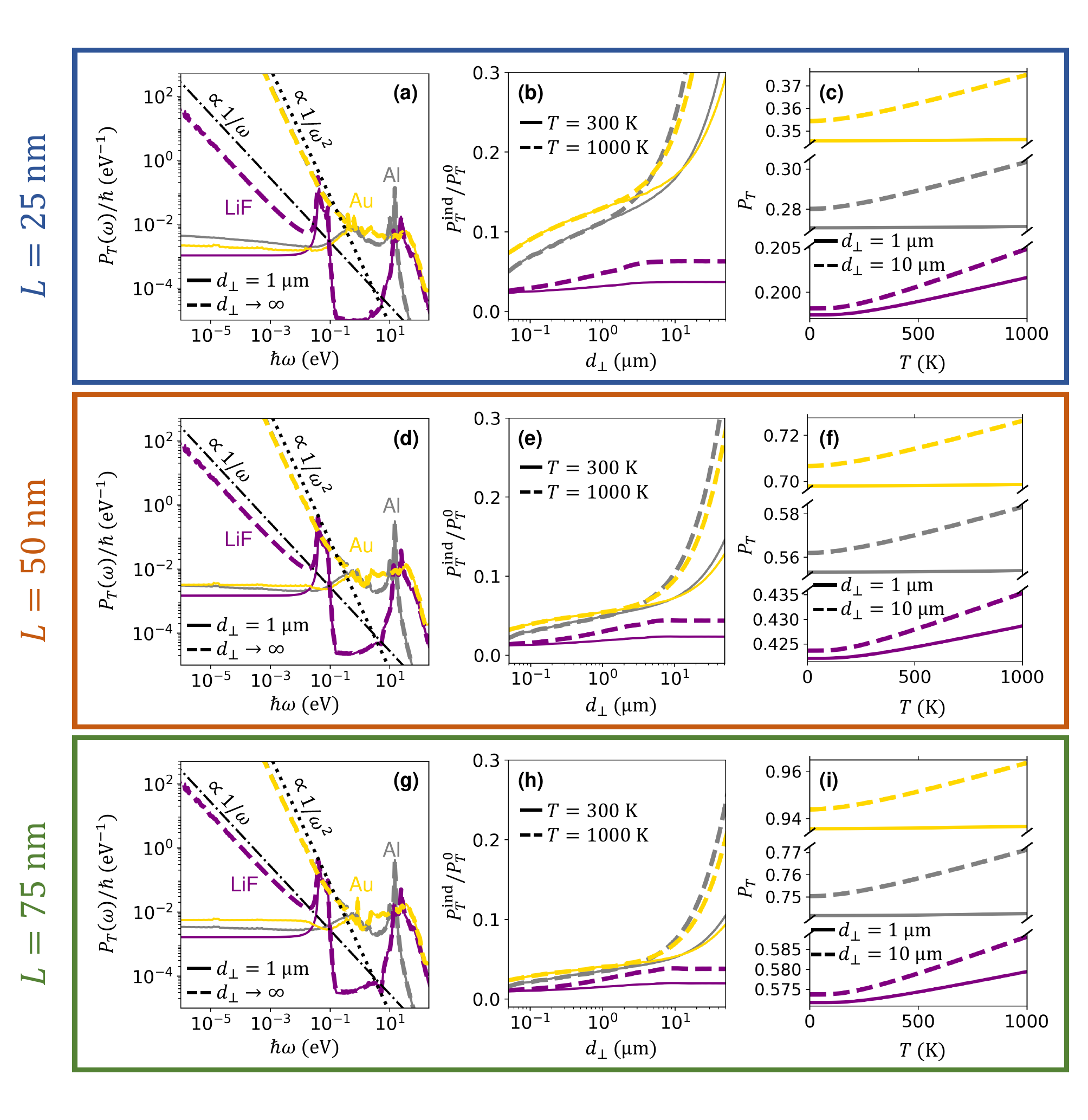}
\caption{\textbf{Effect of film thickness on electron decoherence in the perpendicular e-beam configuration}. \textbf{(a,d,g)}~Frequency-resolved decoherence probability at $T=300$~K for a $200$~keV electron under the configuration of Fig.~4a in the main text when the material is Au, Al, or LiF. We set $d_\perp=1\,\mu$m (solid curves) and $d_\perp\to\infty$ (broken curves, coinciding with the EELS probability). 
\textbf{(b,e,h)}~Ratio $P^{\rm surf}/P^{\rm bulk}$ of the frequency-integrated surface to bulk decoherence probabilities as a function of $d_\perp$ for different values of the temperature.
\textbf{(c,f,i)}~Temperature dependence of the decoherence probability for $d_\perp = 10\,{\rm \mu m}$. The film thickness is $L=25$~nm in (a-c), $L=50$~nm in (b-f), and $L=75$~nm in (g-i).}
\label{FigS4}
\end{figure*}

\end{document}